\documentclass[%
 	superscriptaddress,
 	amsmath,amssymb,
 	aps,
	pra,
    11pt,
    longbibliography
]{revtex4-1}

\usepackage{amsthm}
\usepackage{graphicx}
\usepackage{dcolumn}
\usepackage{bm}
\usepackage{multirow} 
\usepackage{enumerate}
\usepackage{algorithm}
\usepackage{algpseudocode}
\usepackage{tabularx}
\usepackage{hyperref}
\usepackage[mathlines]{lineno}%
\usepackage{float}

\usepackage{subfig}

\newcommand{\ket}[1]{\left|#1 \right\rangle}

\newcommand{\braket}[2]{\left\langle #1| #2 \right\rangle}

\newcommand{\simG}{\textit{sim}($G_1$,$G_2$) }
\newcommand{\CHat}{\hat{C}}
\newcommand{\CHatx}[1]{\hat{C_#1}}
\newcommand{\THat}{\hat{T}}
\newcommand{\nChanges}{\mathcal{C}}
\newcommand{\Erdos}{Erd\H{o}s }
\newcommand{\Renyi}{R\`{e}nyi }
\newcommand{\ErdosRenyi}{Erd\H{o}s-R\`{e}nyi }

\algnewcommand\algorithmicinput{\textbf{Input:}}
\algnewcommand\Input{\item[\algorithmicinput]}
\algnewcommand\algorithmicoutput{\textbf{Output:}}
\algnewcommand\Output{\item[\algorithmicoutput]}
\algnewcommand\algorithmicSubroutine{\textbf{subroutine}}
\algnewcommand\Subroutine{\item[\algorithmicSubroutine]}
\algnewcommand\algorithmicEnd{\textbf{end}}
\algnewcommand\EndSub{\item[\algorithmicEnd]}

\theoremstyle{remark}

\begin{document}

\makeatletter
\newcommand{\rmnum}[1]{\romannumeral #1}
\newcommand{\Rmnum}[1]{\expandafter\@slowromancap\romannumeral #1@}
\makeatother

\title{Quantum walk inspired algorithm for graph similarity and isomorphism}

\author{Callum Schofield}
\affiliation{Department of Physics, The University of Western Australia, Perth, Australia}

\author{Jingbo B. Wang}
\email{jingbo.wang@uwa.edu.au}
\affiliation{Department of Physics, The University of Western Australia, Perth, Australia}

\author{Yuying Li}
\affiliation{Cheriton School of Computer Science, University of Waterloo, Waterloo, Canada}

\date{\today}

\begin{abstract}

Large scale complex systems, such as social networks, electrical power grid, database structure,  consumption pattern or brain connectivity, are often modelled using network graphs. Valuable insight can be gained by measuring similarity between network graphs in order to make quantitative comparisons. Since these networks can be very large, scalability and efficiency of the algorithm are key concerns. More importantly, for graphs with unknown labelling, this graph similarity problem requires exponential time to solve using existing algorithms. In this paper we propose a quantum walk inspired algorithm, which provides a solution to the graph similarity problem without prior knowledge on graph labelling. This algorithm is capable of distinguishing between minor structural differences, such as between strongly regular graphs with the same parameters. The algorithm has a polynomial complexity, scaling with $O(n^9)$.    

\end{abstract}


\maketitle

\section{Introduction}
Many man-made and natural phenomena are modelled using graphs (networks), which show the interconnections between different elements of the systems. These networks have become crucial components in systems which are used in every day life, e.g., Google's web page ranking system or social media networks \cite{socialNetwork}. 
In many applications, it is essential to provide some degree of similarity between two graphs. A number of different applications of graph comparisons have arisen in a wide variety of scientific disciplines. Some examples include classification of objects for the purposes of machine learning \cite{ML}, analysis of protein-protein interaction networks \cite{protein-protein-interaction}, modelling differences in brain connectivity \cite{brainNetwork} and comparing large scale databases \cite{DatabaseComp}. Additionally, these networks may need to be continuously compared for similarities. Examples include identification of faces in images or ensuring computer networks remain secure \cite{DetectionNetwork}. 
	
Algorithms for making inexact comparisons between graphs with known labelling typically have linear or polynomial complexity, such as the \textsc{DeltaCon} algorithm which has, at worst, complexity $O(n^2)$ \cite{DeltaCon}. However, in many applications we do not have the luxury of knowing how to match the nodes (i.e. nodes are not labelled in correspondance). This significantly increases the complexity of comparison algorithms. The unlabelled graph matching problem takes exponential time to compute; additionally the unlabelled sub-graph matching problem is NP-Complete \cite{Bunkegraphmatching,50years}. Due to this, the proposed solutions to the unlabelled graph similarity problem scale exponentially. These methods include the techniques developed to find the maximum common subgraph, such as detecting the maximum clique which has complexity $O((nm)^n)$ \cite{LeviMCS, BunkeMaxComSub} and a decision tree based algorithm with $O(2^nn^3)$ \cite{Bunkegraphmatching}. 
	
This exponential complexity of algorithms for graphs with unknown labelling significantly restricts the size of graphs that can be compared. An improved efficient algorithm must be adopted in order to reliably make comparisons on graphs in today's large data applications. In this paper we propose a technique, developed to utilise coined quantum walks, to make comparisons between undirected graphs. The algorithm is inspired by the work done by Douglas and Wang \cite{DouglasGI, DouglasPhD} which determines whether two input graphs are isomorphic using coined quantum walks. Graph isomorphism can be considered as a binary example of graph similarity, where comparisons output a ``1'' if two graphs are isomorphic, and a ``0'' otherwise. In this paper we extend the graph isomorphism algorithm \cite{DouglasGI, DouglasPhD} to provide a degree of similarity between graphs.  Note that classical graph isomorphic based algorithms have been developed (such as comparing maximum common sub-graphs), however they scale exponentially\cite{50years}. 

The paper is organized as follows. In \S \ref{Sec:graphComp}, we introduce graph comparison problems and discuss the properties required for a graph similarity measure.
The proposed coin quantum walk algorithm for graph similarity measure computation is described in \S\ref{Sec:RW}.
In \S\ref{Sec3: Results}, we present computational results on a variety of graphs to illustrate the proposed method.
Concluding remarks are in \S\ref{Sec:conclude}.
\section{Graph Comprisons and Axioms of Similarity Measures} \label{Sec:graphComp}
\subsection{Graph Comparison}
Graph similarity measures are based on comparisons made between the structure of graphs. Comparison metrics are used to define the distance between graphs or a graph similarity index. Let $G_1(V,E_1)$, $G_2(V,E_2)$ be two graphs defined by a set of vertices $V$ and edge sets $E_i, i=1,2$. Define \simG $\in [0,1]$ as the similarity between these graphs. 
	
Although the concept of similarity between two graphs may be intuitive,  differing similarity measures tend to produce different results due to  differences in how the structures of the graphs are analyzed. In some applications, a binary answer to whether graphs match or not is insufficient, especially if the input graphs are incomplete or contain noise. For graphs which only match approximately, it is useful to be able to determine a measure of similarity. For example, classification techniques in machine learning match objects that are within a certain tolerance level. It is desired that a similarity algorithm returns a score which indicates a degree of how the structures of two networks differ, while also being able to detect slight changes between them.
	
\subsection{Classical Graph Comparison Techniques}
We first review a few existing classical graph comparison techniques.
	
\subsubsection{Maximum Common Subgraph}
Maximum common subgraph techniques involve finding the largest subgraph structure that is common (isomorphic) between the two graphs. If two graphs are isomorphic, their nodes can be relabelled in such a way that they are equivalent ($G_1 \cong G_2$). Similarity metrics that utilise this technique define graphs with larger common subgraphs as being more similar. One such distance metric is defined by Bunke \cite{BunkeMaxComSub}:
\begin{align}
	d(G_1,G_2) = 1 - \frac{\text{n}[\text{mcs}(G_1,G_2)]}{\text{max}(\text{n}[G_1],\text{n}[G_2])},
\end{align}
where n[$G$] denotes the number of nodes in graph $G$, and mcs($G_1,G_2$) is the maximum common subgraph of $G_1$ and $G_2$. Note that as this metric defines the distance between graphs a smaller value would denote a higher similarity. 
	
\begin{figure}[h]
    \centering
    \subfloat[]{\includegraphics[width=0.25\textwidth]{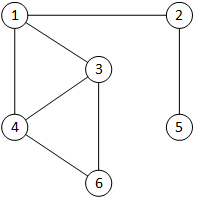}}
	\subfloat[]{\includegraphics[width=0.25\textwidth]{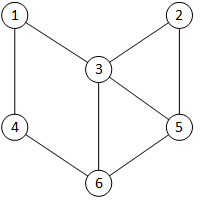}}
    \subfloat[]{\includegraphics[width=0.155\textwidth]{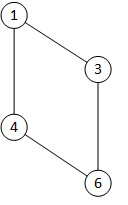}}
    \caption{ Example 6 node graphs (a) and (b). (c) shows the maximum common subgraph of (a) and (b).}
    \label{Fig 1-Ex}
\end{figure}

Figure \ref{Fig 1-Ex} gives an example maximum common subgraph of two graphs. In this example, the distance would be
\begin{align}
	d(G_1,G_2) = 1 - \frac{4}{6} = 0.33.
\end{align}

\subsubsection{Graph Edit Distance}
The graph edit distance between two graphs is defined as the number of changes that would be required to make two graphs the same. These changes include relabelling vertices and adding or removing vertices and edges \cite{Bunkegraphmatching}. Additionally, each action can be attributed a weighting to provide variation to the resulting similarity measure. For example, a minimum of four changes would be required to make graph (a) in Figure \ref{Fig 1-Ex} equivalent to graph (b). 
	
\subsubsection{Graph Eigensystems}
The eigensystems of graphs can also be utilised to define the distance between two graphs. For example, consider two undirected graphs $G_1$ and $G_2$ with adjacency matrices $A$ and $B$ respectively. Umemyama \cite{Umeyama_Eigen} defines the shortest distance between $G_1$ and $G_2$ by minimising:
\begin{align}
	J(P) &= \lVert P A P^T - B \rVert^2,
\end{align}
where $P$ is a permutation matrix which represents a map which relates the nodes in $G_1$ to those in $G_2$. This permutation is  optimal for the shortest distance, which is found by using the eigen decompositions of $A$ and $B$. 

\subsubsection{\textsc{DeltaCon}}
\textsc{DeltaCon} is a graph similarity algorithm developed by Koutra, Vogelstein and Faloutsos \cite{DeltaCon}. The algorithm makes comparisons on the pairwise node affinities of graphs. Due to these pairwise comparisons, the graph must have known labelling. A matrix $S$ is constructed, using Equation \eqref{Eq1: S_NA}, such that $S_{ij}$ indicates the influence of node $i$ on node $j$: 
\begin{align}
    S &= \left( I + \epsilon^2D -\epsilon A \right)^{-1},
    \label{Eq1: S_NA}\\
    \epsilon &= \frac{1}{1 + \text{max}(D)},
\end{align}
where $I$ is the identity matrix, $D$ is the diagonal degree matrix ($d_{ii}$ is the degree of node $i$) and $A$ is the adjacency matrix. The distance and similarity between two graphs is then defined by
\begin{align}
    d &= \sqrt{\sum_{i=1}^{n}\sum_{j=1}^{n}\left(\sqrt{S_{1,ij}}-\sqrt{S_{2,ij}}\right)^2},\\
    sim(G_1,G_2) &= \frac{1}{1 + d} \label{EQ1: S_sim},
\end{align}
where $d$ is the Matusita distance between the node affinities. This distance is used in order to boost the graph node affinities, which increases the sensitivity of the metric. The resulting similarity score is in the interval $[0,1]$. 

\subsection{Axioms}
\newcounter{Axiom}
\renewcommand{\theAxiom}{A\arabic{Axiom}}
\newcounter{Property}
\renewcommand{\theProperty}{P\arabic{Property}}
The given distance between a pair of graphs relies on the similarity method used. It is not necessarily the case that two metrics will return the same similarity score. The requirements of the similarity metric would also differ between applications. However, Koutra et al. have formalised a set of axioms that graph similarity measures should conform to \cite{DeltaCon}:


\label{Sec: Axioms}
\begin{itemize}
\refstepcounter{Axiom}\label{A1}
\item [A1] Identity: \simG = 1 $\Leftrightarrow G_1 \cong G_2$.
\refstepcounter{Axiom}\label{A2}
\item [A2] Symmetric: \simG = \textit{sim}($G_2$,$G_1$).
\refstepcounter{Axiom}\label{A3}
\item [A3] Zero: For a complete graph $G_1$ and empty graph $G_2$, \simG $\rightarrow 0$ as $n \rightarrow \infty$.
\end{itemize}


Another important requirement raised by Koutra et al. is that the algorithm used must be scalable \cite{DeltaCon}. Many applications require lots of comparisons to be made on large graphs, thus the similarity algorithm must be able to compute comparisons efficiently on large sets.

\subsection{Graph Representations}
A graph, denoted by $G(V,E)$, consists of a set of vertices (or nodes) $V$, representing position states, and a set of edges $E$, denoting connections between vertices. A graph can be described via a $n\times n$ adjacency matrix $A$:  
\begin{align}
  A_{ij} &= 
  \begin{cases}
    1, \qquad  \text{if edge } \{i,j\}\in E\\
    0, \qquad  \text{otherwise}.
  \end{cases}	
\end{align}
Note that for undirected graphs, the adjacency matrix is symmetric ($A_{ij}=A_{ji}$). 
More generally, a real valued weight $A_{ij}$ can be assigned to the edge $\{i,j\}\in E$. The degree  $d_i$ of the node $i$, in an undirected graph is,
\begin{align}
  d_i = \sum_{j=1}^{n}A_{ji}.
\end{align}
Since  the unlabelled graph matching problem is NP-Complete \cite{Bunkegraphmatching,50years},  existing solutions to measure graph similarity would take exponential time to compute. Next we propose a coined quantum walk solution.

\section{Coined Quantum Walk Graph Similarity} \label{Sec:RW}

\subsection{Coined Quantum Walk}
Classical random walks, which are stochastic processes, can describe the path of a random walker on a graph. The quantum analogue, quantum walks, describe the time evolution of the probability state of a walker on a graph. Both discrete and continuous time formulations of walks exist, however we will restrict the discussion to the former.

At each time step, a classical walker flips a ``coin'' to determine which direction to move in next. Similarly, every step in a quantum walk involves applying a coin operator $\CHat$, then translating the system according to the coin state via the translation operator $\THat$ \cite{JWQuantumWalks}. 

Consider an arbitrary graph described by the adjacency matrix $A$. The coined quantum walk formalism is restricted to undirected graphs. For a graph with $n$ nodes, there are $n$ position states labelled $\ket{v_x} \equiv \ket{x}$. The degree of each position state $\ket{x}$ gives the number of coin states $\ket{c}$. This is equivalent to a walker having $d_x$ paths to choose at position $\ket{x}$. Each node has a local coin operator $\CHatx{x}$, represented by a $d_x \times d_x$ matrix. 

As the algorithm must work for graphs with unknown labelling, the coin operator used must not have a bias with respect to the neighboring nodes. This means that the coin operator must be constructed such that it will produce the same result for all combinations of labelled nodes. This constrains each coin to be in the form \cite{DouglasGI}:
\begin{align}
	\CHat_{ij} &= 
	\begin{cases}
		a, \qquad  i = j\\
		b, \qquad i \neq j,
	\end{cases}
\end{align} 
where $a,b\in \mathbb{C}$. One such unitary operator which keeps the symmetry with respect to its neighboring nodes is the Grover coin:
\begin{align}
	\CHat_{ij} &= \frac{2}{d} - \delta_{ij},
\end{align} 
where $d$ is the degree of the corresponding node and $\delta_{ij}$ is the Kronecker delta function.

Considering an edge $e_{xy}$ connecting nodes $x$ and $y$, one formalism of the 1-D quantum walk to a 2-D graph defines the translation operator as the unique mapping from one state to another along an edge:  
\begin{align}
	\THat \ket{x,d_{x,i}} \longmapsto \ket{y,d_{y,j}}, 
\end{align}
where $d_{x,i}$ is the $i$th neighbour of $x$. To achieve this, Watrous proposed to label the coin states of each node via the edges connected to it \cite{Watrous_QW}. This allows the system to be represented by the set of nodes $x$ and edges connecting pairs of nodes $e_{xy}$. The state of the system is then represented in general as 
\begin{align}
	\ket{\psi} &= \sum_{x=1}^{n} \sum_{y}\alpha_{xy}\ket{x,e_{xy}},
\end{align}
where $\alpha_{xy}$ denotes the probability amplitude of the walker at node $x$ transitioning to $y$, which satisfies the normalization condition such that  $\sum_{x,y}|\alpha_{xy}|^2=1$.
The translation operator is then defined by shifting the walker along an edge:
\begin{align}
	\THat \ket{x,e_{xy}} = \ket{y,e_{yx}}.
	\label{Eq3: EdgeTrans}
\end{align}
The quantum walk is then carried out via repeated application of the coin and then translation operator, such that after $t$ time steps the state of the system is given by: 
\begin{align}
	\ket{\psi_t} &= \left(\THat\CHat\right)^t\ket{\psi_0}.
\end{align}
After observation, the probability the walker is at node $x$ is given by:
\begin{align}
	P_t(x) &= \sum_{y} |\braket{x,e_{xy}|\psi_t}|^2. \label{Eq3: Pn}
\end{align}

Not knowing the node labelling also requires us to ensure that there is no bias between nodes. To do this, the initial state is constructed in equal superposition, as in equation \eqref{Eq3: psi0}. For simplicity, the coin states $\ket{e_{xy}}$ are denoted as $\ket{e_{xj}}$, where $j = 1,\ldots, d_x$, and $\ket{e_{xj}}$ is the $j$th outgoing edge of vertex $x$. Thus
\begin{align}
	\label{Eq3: psi0}
	\ket{\psi_0} &= \sum_{x=1}^{n} \sum_{y}\frac{1}{\sqrt{n d_x}}\ket{x,e_{xy}}
	= \sum_{x=1}^{n} \sum_{j=1}^{d_x}\frac{1}{\sqrt{n d_x}}\ket{x,e_{xj}}. 
\end{align}

\subsection{Graph Comparison Score}
The graph isomorphism algorithm developed by  Douglas and Wang \cite{DouglasGI} generates graph comparison scores, an integer value that can be used to make comparisons between graphs. These comparison scores are calculated by comparing the time evolution of quantum walks on the graphs. In some cases, such as strongly regular graphs with the same parameters, a simple quantum walk is not sufficient to distinguish between their structure. To alleviate this, a phase is added to a number of reference nodes in each graph, which produces a new level of interaction in the walk \cite{DouglasGI}. This interaction helps break symmetries in the graph.

\begin{figure}[t]
	\centering
	\includegraphics[width=0.8\textwidth]{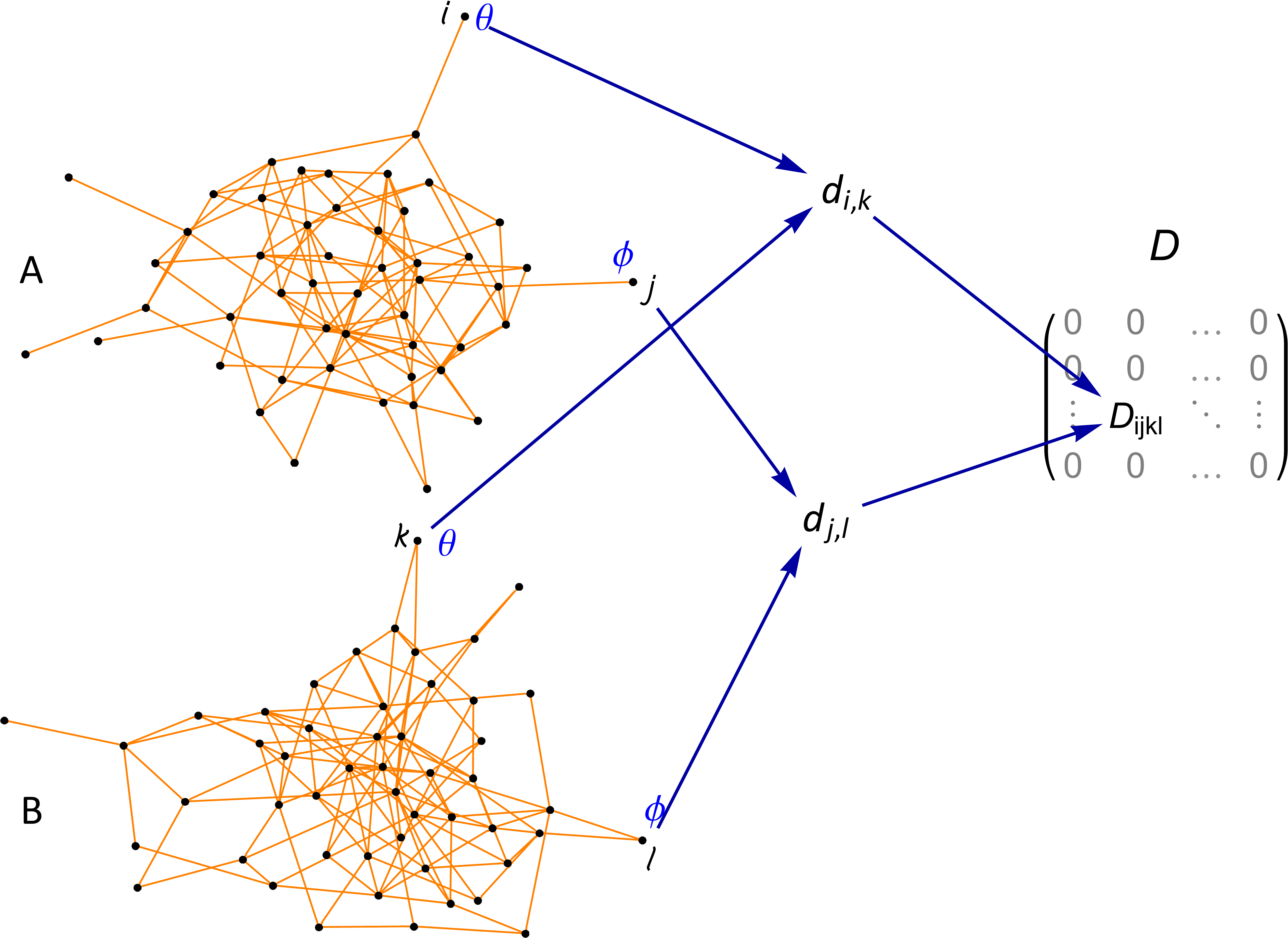}
	\caption{Schematic to obtain $D_{ijkl} = d_{ik} + d_{jl}$, with reference nodes $\{i,j,k,l\}$.}	
	\label{Fig: Schematic}	
\end{figure}

We adopt a two-node phase addition scheme, chosen as it is the fewest number of reference nodes required to distinguish SRGs \cite{DouglasGI}. Consider two input graphs, $A$ and $B$. Let $i,j\in A$ and $k,l\in B$ be the chosen reference nodes, as in Figure \ref{Fig: Schematic}. Denote two phases $\phi, \theta$. At each step of the walk these phases are applied to the reference nodes:

\begin{minipage}{0.5\textwidth}
	\begin{align*}
	\ket{i,c} &\longmapsto e^{i\theta}\ket{i,c},\\
	\ket{j,c} &\longmapsto e^{i\phi}\ket{j,c},
	\end{align*}
\end{minipage}%
\begin{minipage}{0.5\textwidth}
	\begin{align*}
	\ket{k,c} &\longmapsto e^{i\theta}\ket{k,c},\\
	\ket{l,c} &\longmapsto e^{i\phi}\ket{l,c}.
	\end{align*}
\end{minipage}\vspace{0.5em}
The phase values chosen must be in the range $(0,2\pi)$, with $\phi\ne\theta$. The phase values should also be sufficiently different to ensure that each reference node is distinguishable. The similarity measure result is also not entirely independent of the chosen phase values. However, for sensible values of phase the influence is minimal, generally resulting in a range of at worst $\pm 0.05$ in similarity score. Additionally, this is not an issue if the particular application only requires to know the relative similarity between sets of graphs, assuming the phase is kept consistent. 


A graph comparison 4-D matrix is constructed by making comparisons between the probability amplitudes of each pair of reference nodes at every step of the walk. To ensure that the complete structure of the graph is explored by the walker, it must complete at least twice as many steps as the diameter of the graph, which is the maximum shortest path between all pairs of nodes in the graph. The graph comparison score is then defined as the sum of the elements in $D$, where 
\vspace{-1em}
\begin{align}
	D_{ijkl} &= d_{ik} + d_{jl},
\end{align}
and $d_{xy}$ is the distance between the probability amplitude sets of reference nodes $x \in A$ and $y \in B$. For testing purposes a number of different distance measures have been tried (see \S \ref{Sec: Distance Metric} for details). 


As the labelling of each graph is not known, the probability amplitudes need to be compared for all unique combinations of $\{i,j,k,l\}$. If the two nodes match up, or are similar, they are given a low score. The full comparison matrix has $n^4$ elements, however this can be reduced due to the symmetry in the adjacency matrices. As we are applying the same phase to each pair of reference nodes, and the difference in the response is independent of the phase applied, testing combinations of $\{j,i\}$ will give the same results as $\{i,j\}$. Thus, the number of comparisons necessary can be reduced to $\dfrac{n^3(n-1)}{2}$.

\begin{figure}[H]
	\centering
    \subfloat[]{\includegraphics[width=0.45\textwidth]{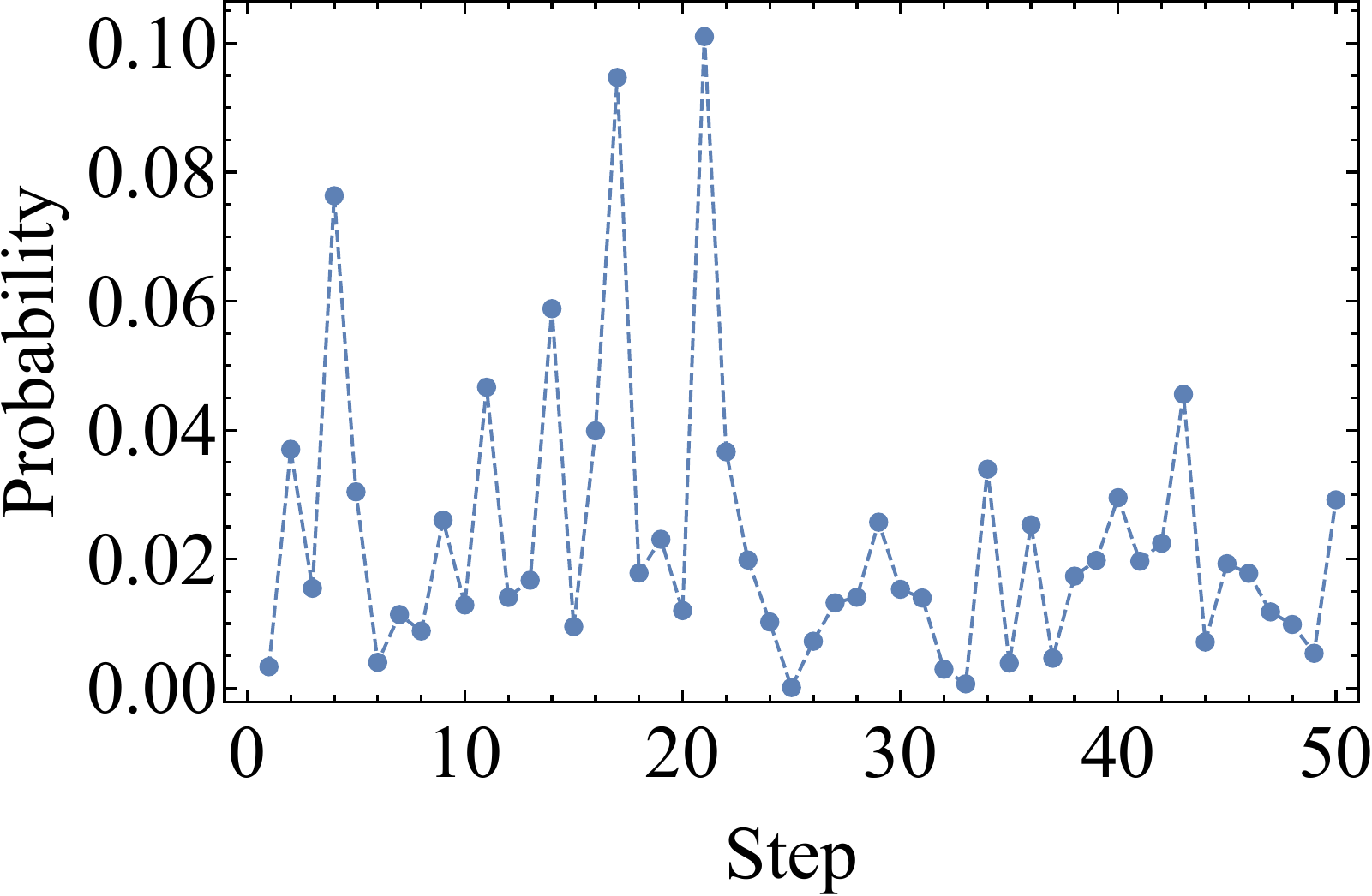}}\qquad
    \subfloat[]{\includegraphics[width=0.45\textwidth]{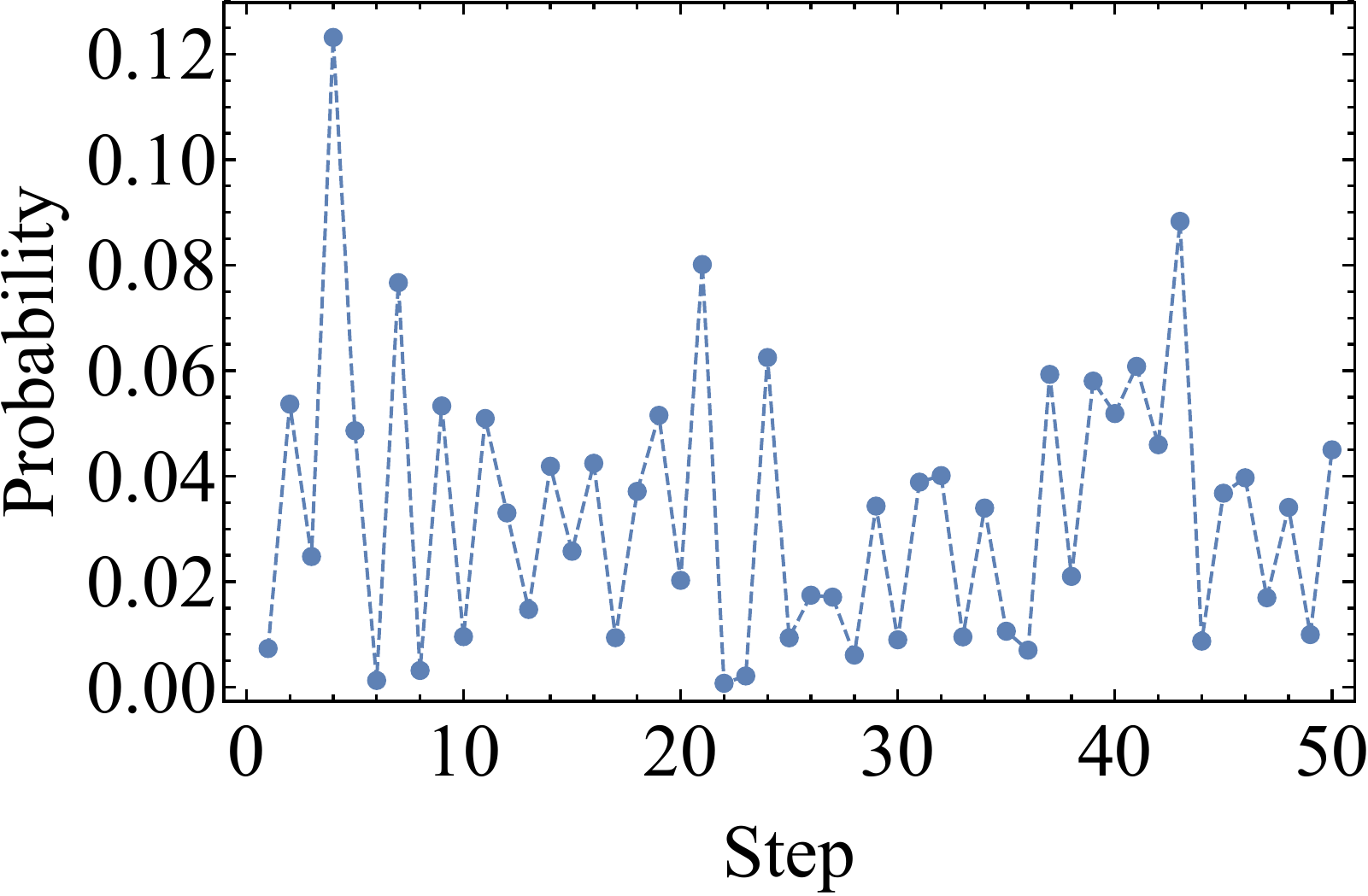}}\\
    \subfloat[]{\includegraphics[width=0.45\textwidth]{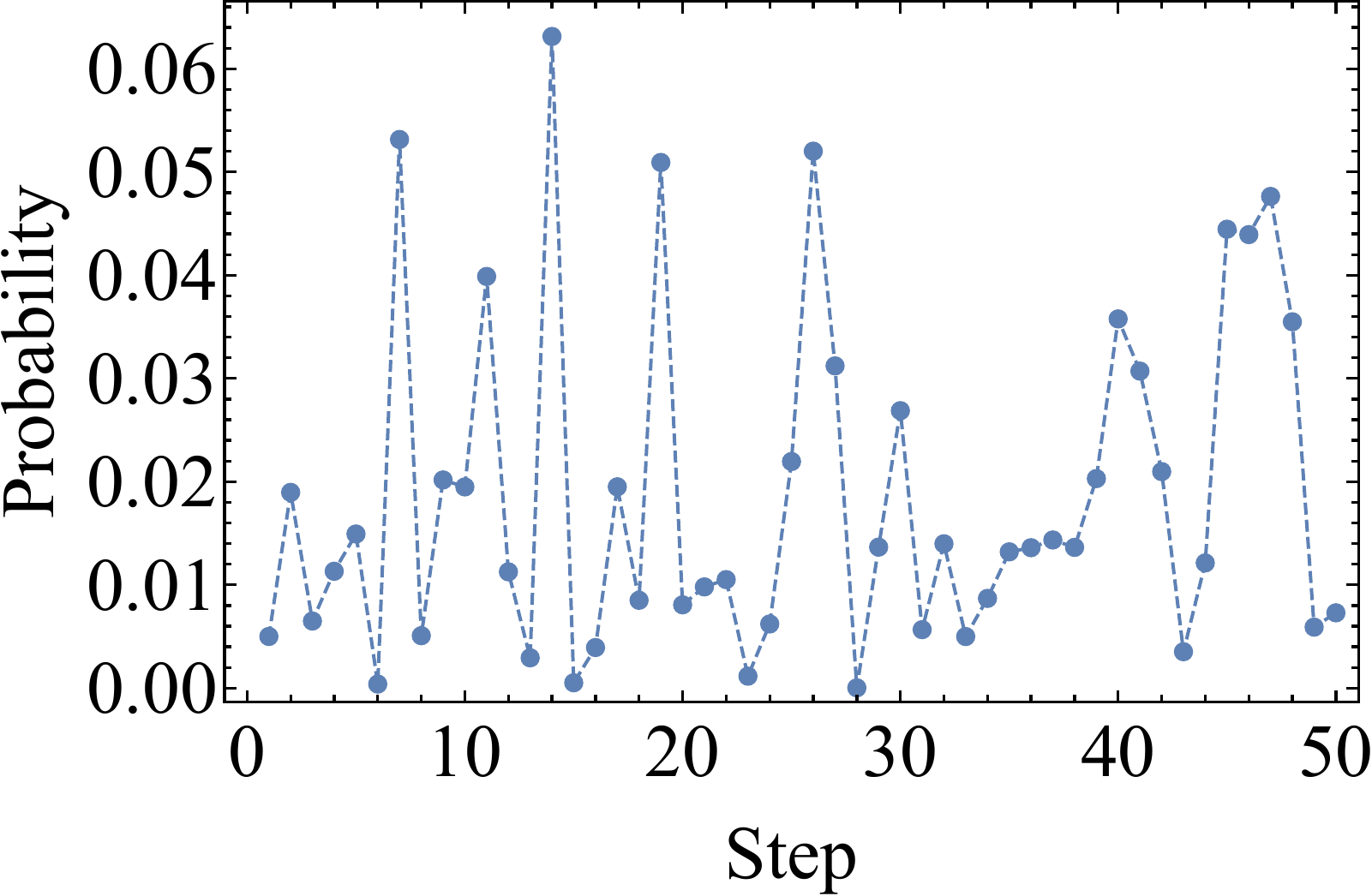}}\qquad
    \subfloat[]{\includegraphics[width=0.45\textwidth]{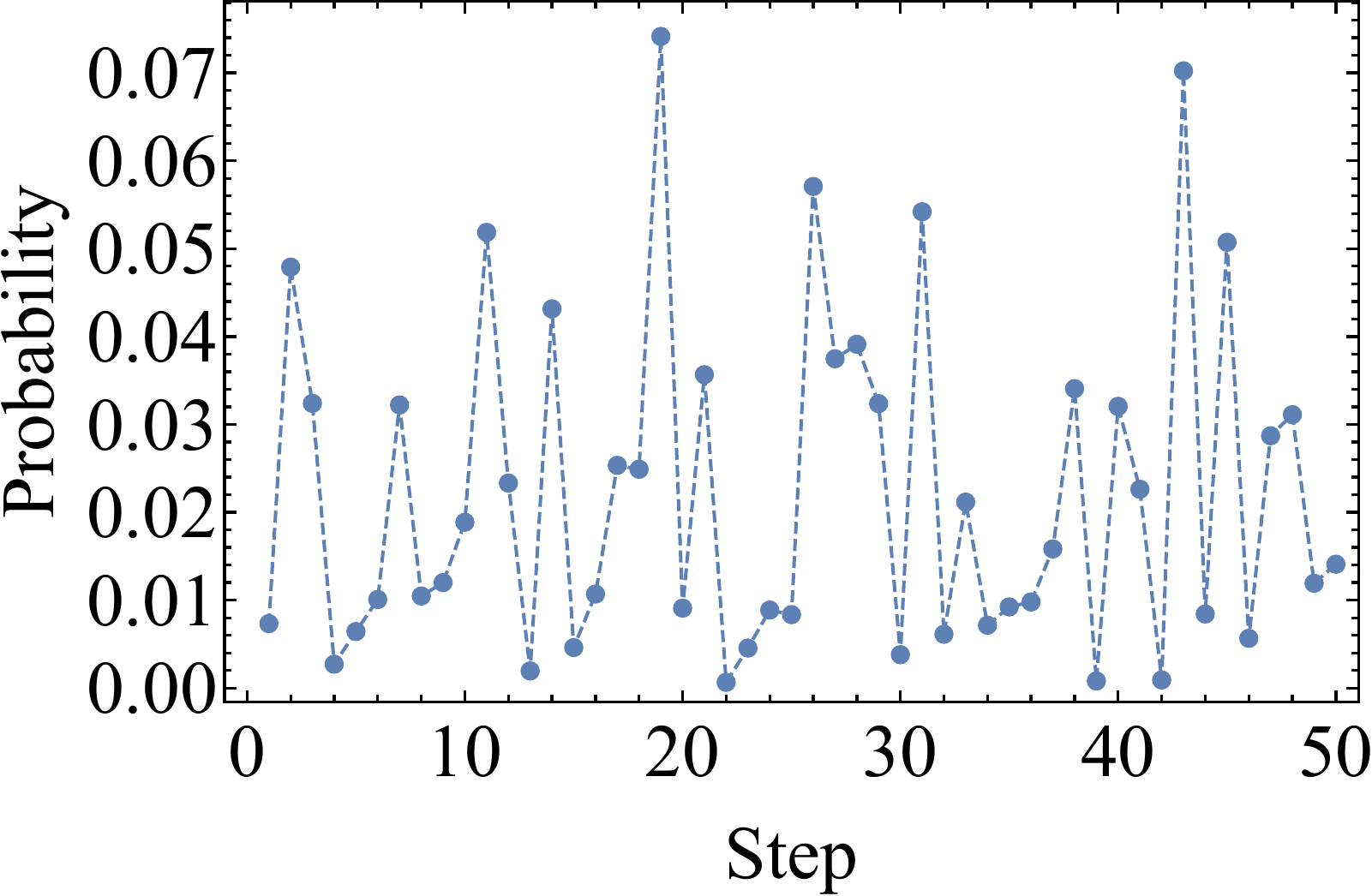}}
	\caption{Probability time evolution of a selected combination of reference nodes $\{i,j,k,l\}$  in Fig. \ref{Fig: Schematic} with applied random phases $\theta$ and $\phi$. The comparisons $d_{ik}$ and $d_{jl}$ are defined between probability vectors $prob_i$ $\leftrightarrow$ $prob_k$ and $prob_j$ $\leftrightarrow$ $prob_l$ respectively. The comparison score is determined by the sum of all unique combinations $D_{ijkl} = d_{ik} + d_{jl}$.}
	\label{Fig: GraphScore}
\end{figure}

\subsection{Distance Metric}
\label{Sec: Distance Metric}
We compare the probability amplitudes at the reference nodes  using a variety of distance measures. In the following, we denote the set of probability amplitudes of reference node $i$ as $prob_i$ with input graphs $G_1$ and $G_2$, where $\{i,j\}\in G_1$ and $\{k,l\}\in G_2$. The first measure is used by Douglas and Wang in their graph isomorphism algorithm \cite{DouglasGI}.

\begin{enumerate}
	\item Euclidean distance with $\epsilon$ threshold: 
	\begin{align*}
	D_{ijkl} &=
	\begin{cases}
	0, \quad  \text{if }~ \lVert prob_i - prob_k \rVert + \lVert prob_j - prob_l \rVert  < \epsilon\\
	1, \quad \text{otherwise.}
	\end{cases}
	\end{align*}
\end{enumerate}
The thresholding method is useful in the graph isomorphism case, where we require a sensitive metric to determine whether two graphs are isomorphic. However, graph similarity is not a binary problem and  we lose significant information regarding the differences between the two graphs when thresholding the results. Thus, removing  thresholding from the graph isomorphism algorithm would likely improve the graph similarity performance (while possibly reducing its effectiveness at determining graph isomorphism for extreme cases such as SRGs).  In addition to the Euclidean distance, two other metrics have been tried in an attempt to increase the sensitivity of the algorithm. The Euclidean distance squared metric boosts the distance between each pair of probability vectors by squaring the Euclidean distance between them. The Matusita distance metric boosts the probability vectors before the comparison is made. The precise definitions for these three metrics are given below.
\begin{enumerate}
    \setcounter{enumi}{1}
	\item Euclidean distance: 
	\begin{align*}
	D_{ijkl} = \frac{\lVert prob_i - prob_k \rVert + \lVert prob_j - prob_l \rVert}{nSteps}.
	\end{align*}
	\item Euclidean distance squared: 
	\begin{align*}
	D_{ijkl} = \left(\frac{\lVert prob_i - prob_k \rVert}{nSteps}\right)^2 + \left(\frac{\lVert prob_j - prob_l \rVert}{nSteps}\right)^2.
	\end{align*}
	\item Matusita distance: 
	\begin{align*}
	D_{ijkl} = \frac{\sqrt{\sum_{m=1}^{n}\left(\sqrt{prob_i(m)}-\sqrt{prob_k(m)}\right)^2}+\sqrt{\sum_{m=1}^{n}\left(\sqrt{prob_j(m)}-\sqrt{prob_l(m)}\right)^2}}{nSteps}.
	\end{align*}
\end{enumerate}

As the sum of the probabilities at each time step must equal one, the Euclidean difference between the probability amplitude of two nodes would be in the range $[0,1]$. A value of $0$ would denote the probability of finding the walker at both reference nodes as being the same at all time steps. Thus, the distance between the two sets of probability vectors corresponding to a pair of reference nodes would be in the range $[0,nSteps]$. As we do not want the similarity metric to be dependent on the number of steps the walker takes, the distance is normalised with respect to $nSteps$. Since $D_{ijkl}$ is given by the sum of the distance between two pairs of reference nodes, it is in the range $[0,2]$.

\subsection{Similarity Metric}
Consider the comparison between two input graphs, $A$ and $B$, with a graph comparison score $D_{AB}$. This score will be finite and non-zero, regardless of the input graphs. In order to be consistent with the identity axiom (\ref{A1}), we need to compare $D_{AB}$ with the resulting comparison score assuming $A$ was isomorphic to $B$. This is determined by calculating the comparison score between $A$ and an isomorphic permutation of $A$ (denote this $A'$, and the resulting comparison score $D_{AA'}$). Similarly the comparison score $D_{BB'}$ is calculated. If $D_{AB}$ is equivalent to both $D_{AA'}$ and $D_{BB'}$, the input graphs are said to be isomorphic \cite{DouglasGI}.

The distance between the input graphs ($A$ and $B$) is then defined to be the normalised difference between $D_{AB}$ and the corresponding isomorphic comparison scores, i.e.:
\begin{align}
  \label{Eq: SimDistance}
  \mathcal{D} &= \frac{\bigl| D_{AA'}-D_{AB}\bigr|}{D_{AA'}} + \frac{ \bigl| D_{BB'}-D_{AB}\bigr|}{D_{BB'}},
\end{align}
while the similarity metric is given by:
\begin{align}
  \label{Eq: SimMetric}
  sim(A,B)&=\frac{1}{1+\mathcal{D}}.
\end{align}

\begin{figure}[t]
	\centering
	\includegraphics[width=0.85\textwidth]{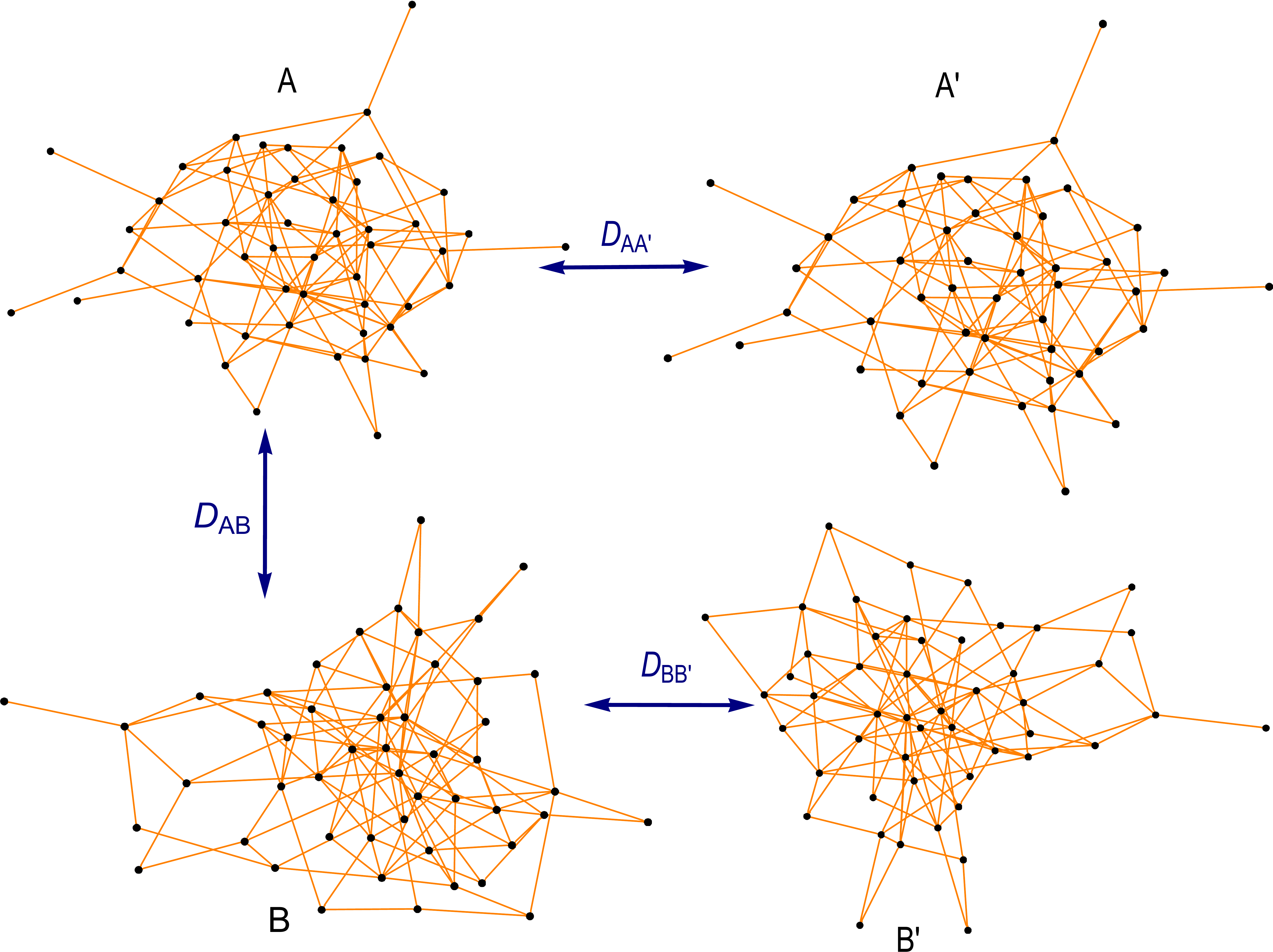}
	\caption{Comparison scores $D_{AB},D_{AA'},D_{BB'}$ are calculated between input graphs $A,B$ and isomorphic copies $A',B'$.}
	\label{Fig: QWMetric}
\end{figure}

We summarize the proposed coin quantum walk graph similarity measure computation in Algorithm 1.

\begin{algorithm}[H]
	\caption{Graph similarity measure}
	\label{Alg: GI}
	\begin{algorithmic}
 		\Input 
        \State $A \gets \text{Adjacency matrix for graph }A$
        \State $B \gets \text{Adjacency matrix for graph }B$
        \State $nSteps \gets \text{Number of steps the quantum walker completes}$
		\Output
        \State $sim \to$ Graph similarity score
		\Function{GraphSim}{$A,B,nSteps$}
        	\State $A' \gets$ isomorphic permutation of $A$ 
            \State $B' \gets$ isomorphic permutation of $B$ 
            \State $D_{AA'} \gets$ \Call{ComparisonScore}{$A,A',nSteps$}
          	\State $D_{BB'} \gets$ \Call{ComparisonScore}{$B,B',nSteps$}
          	\State $D_{AB} \gets$ \Call{ComparisonScore}{$A,B,nSteps$}
            \State $\mathcal{D} = \frac{\bigl| D_{AA'}-D_{AB}\bigr|}{D_{AA'}} + \frac{ \bigl| D_{BB'}-D_{AB}\bigr|}{D_{BB'}}$
        	\State\Return $sim = \frac{1}{1+\mathcal{D}}$
        \EndFunction
        \Statex
        \Function{ComparisonScore}{$A,B,nSteps$}
			\State Initialise states $\ket{\psi}_A, \ket{\psi}_B$ in equal superposition and generate each local coin, $\CHatx{v}$ \Comment{v=1,\ldots,n}
            \State $D = 0$
			\For{$i = 1,n-1$}
				\For{$j = i+1,n$} 
  					\State $prob_A \gets$ Propagate $\ket{\psi}_A$  for $nSteps$ to find probability time evolution at nodes $i$ and $j$.
					\For{$k = 1,n$}
						\For{$l = 1,n$} 
							\State $prob_B \gets$ Propagate $\ket{\psi}_B$  for $nSteps$ to find probability time evolution at nodes $k$ and $l$. 
							\State $d_1 \gets $ Distance between nodes $i$ and $k$
							\State $d_2 \gets $ Distance between nodes $j$ and $l$
							\State $d = \dfrac{d_1+d_2}{nSteps}$
							\State $D = D + d$
						\EndFor	
					\EndFor
				\EndFor	
			\EndFor
			\State\Return D
		\EndFunction
	\end{algorithmic}
\end{algorithm}

\subsection{Algorithm Complexity}
\label{Sec: complex}

Consider an undirected graph with $n$ nodes and $k$ edges, where $k \le n^2/2$ (allowing for self loops). Let $d_i$ denote the degree of node $i$. In the worst case, a complete graph, $d_i = n ~ \forall ~ i$. Each step of the walk involves a coin flip and translation operation. The coin operator, $\CHatx{i}$, is applied locally to each node. Each local coin is of size $(d_i\times d_i)$ and is applied to the coin states of node $i$, which is a vector of length $d_i$. Note that there are $2k$ coin states, with $\sum_{i}d_i = 2k$ (in the worst case, $2k=n^2$). The application of the local coin operator then scales with $O({d_i}^3)\le O(n^3)$. As this operator is applied $n$ times, the total coin flip scales with $O(\sum_{i}^{n}{d_i}^3) \le O(n^4) = O(k^2)$.

The translation operator, $\THat$, is applied by shifting states along edges using a temporary register, thus it only scales with $O(n^2)$. Therefore, each step of the quantum walk is $O(n^4+n^2)=O(n^4)$.  During each step, there is an additional calculation to map the probability amplitudes of each state to the probability of the walker being at each node in the graph. This however only scales with $O(n)$, which is dwarfed by the scaling for the quantum walk operations.

As the walker needs to complete at least as many steps as twice the diameter of the graph, the walker must make at most $2n$ steps. Thus the walk would scale  $O(2\mathcal{D}n^4)\le O(n^5)$, where $\mathcal{D}$ is the diameter of the graph. Since the diameter of a graph is given by the maximum shortest path between all pairs of nodes, this tends to have a large variation between different graph structures. For each combination of nodes $\{i,j,k,l\}$ tested, two walks are conducted and the distance between the probability vectors for the two pairs of nodes is calculated. This scales at worst with $O(2n^5 + 2n)=O(n^5)$. To calculate each graph comparison score $\dfrac{n^3(n-1)}{2}$ combinations of reference nodes must be tested, thus the algorithm which calculates three comparison scores scales with $O\left(\dfrac{3n^8(n-1)}{2}\right)=O(n^9)$. Thus, assuming that each operation has $O(1)$ complexity, this algorithm has polynomial time complexity. 

\section{Results}
\label{Sec3: Results}

\subsection{Sample Random Graphs}
A variety of different random graph models have been utilised to test the algorithm against the graph similarity axioms and properties. Additionally, this will determine whether the algorithm has a bias towards certain graph structures. All random graphs used for testing have been generated using \textit{Mathematica}. Graphs referred to as ``Random graphs'' are generated using a uniform distribution (the probability density function of vertex degree is uniform). These graphs will be denoted as $R(n,e)$, where $n$ is the number of nodes and $e$ is the number of edges in the graph. Other sets of graphs used to test the graph similarity algorithms are detailed below.

\subsubsection{\ErdosRenyi Graphs}
Random graph theory was pioneered by P. \Erdos and A. \Renyi in 1959. \ErdosRenyi graphs are constructed by choosing whether or not to add an edge connecting every pair of nodes using probability $p$ \cite{ER_RandomGraphs,ER_Evolution}. The choice for each edge is independent to all other connections. \ErdosRenyi random graphs will be denoted using $ER(n,p)$ where $n$ is the number of nodes and $p$ is the probability that any two nodes are connected. These graphs have been used in practice to describe and analyse complex topological networks, such as modelling epidemics as a dynamical computational networks \cite{ER_Use_Epidemic}.

\subsubsection{Scale-Free Graphs}
Albert et al. first discovered the scale-free nature of networks which model the topology of the world-wide web (WWW) in 1999. These networks follow a power law degree distribution where the probability $P(k)$ of a node connecting with $k$ other nodes scales with $P(k) \sim k^{-\gamma}$ for some constant $\gamma$ \cite{SF_EmergenceScaling,SF_WWW}. Such networks maintain this power law distribution regardless of the number of nodes. This scale-free property was subsequently found in other networks describing real world systems, such as cellular protein-protein interactions \cite{SF_protein-protein1,SF_protein-protein2}, brain activity \cite{SF_BrainActivity}, and interaction in social networks \cite{SF_Social}.

Scale-free networks are constructed by starting with a small graph and repeatedly adding nodes. Each new node added is connected to $m$ other nodes already in the network. This scheme provides preferential attachment, as nodes will be more likely connected to nodes with high connectivity \cite{SF_Characteristics}. Scale-free graphs will be denoted $SF(n, m)$, where each new node added is connected to $m$ other nodes and the total number of nodes in the network is given by $n$.

\subsubsection{Strongly Regular Graphs}
A $k$-regular graph with $n$ nodes is said to be strongly regular if all adjacent nodes have $\lambda$ common neighbours and all non-adjacent nodes have $\mu$ common neighbours \cite{SRG}. Due to this structure, it is difficult to distinguish between SRGs with the same parameters ($n,k,\lambda,\mu$). This provides a good benchmark to test for isomorphism to ensure a similarity measure correctly distinguishes between similar graph structures. Strongly regular graphs with parameters ($n,k,\lambda,\mu$) will be denoted $SRG(n,k,\lambda,\mu)$.

\subsection{Isomorphism and Identity Testing}
\label{Sec3: Isomorphism Testing}
The identity axiom (\ref{A1}) is a key property of the graph similarity metric. It is expected that the algorithm will output ``1'' when the two input graphs are the same. As mentioned above, SRGs with the same parameters ($n,k,\lambda,\mu$) are difficult to distinguish, which make them a good baseline for testing the algorithm. These tests are used to confirm the similarity measure can correctly distinguish whether a pair of graphs differ in any way. 

\begin{figure}[h]
	\centering
    \subfloat[]{\includegraphics[width=0.23\textwidth]{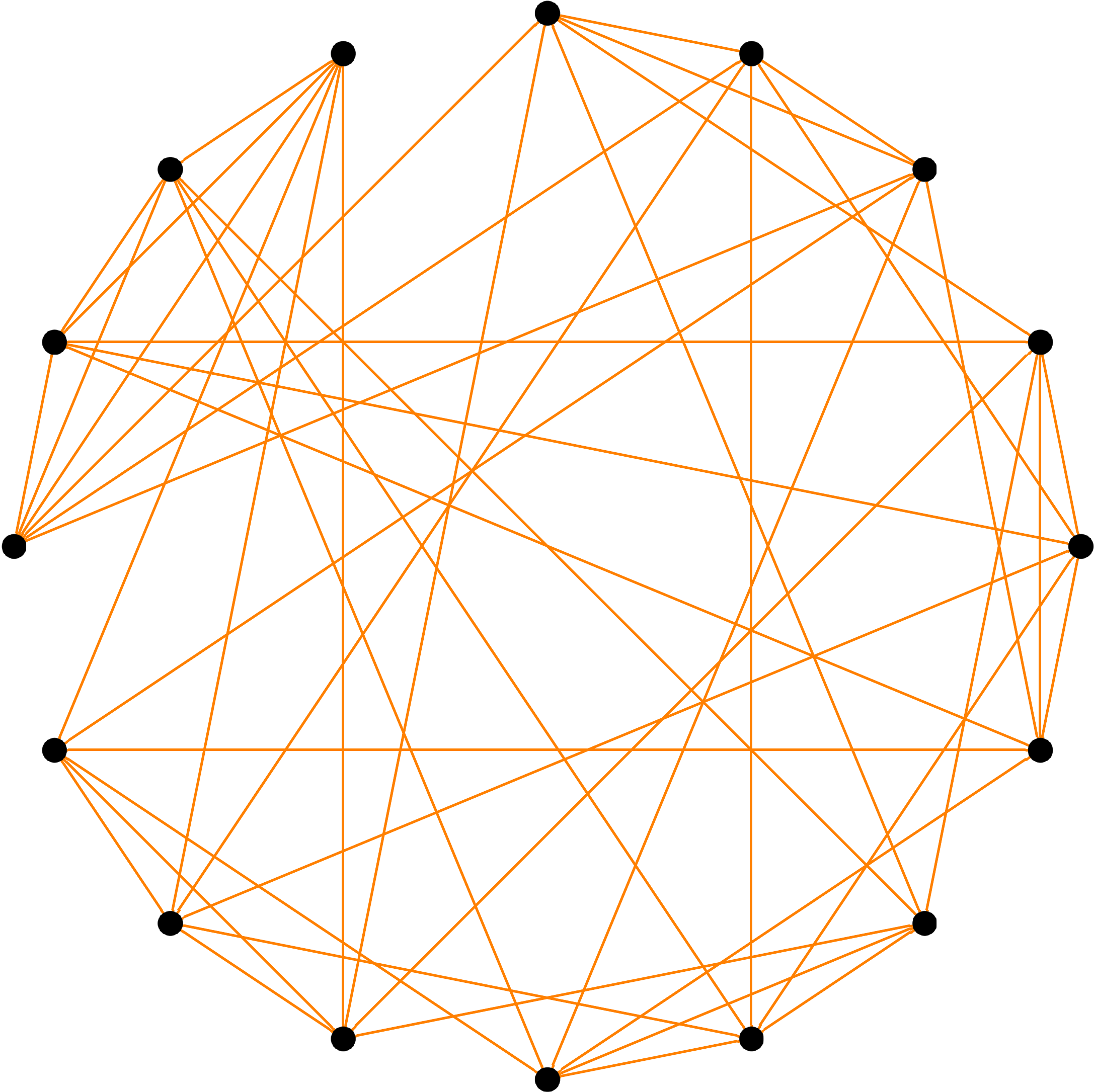}}\;
    \subfloat[]{\includegraphics[width=0.23\textwidth]{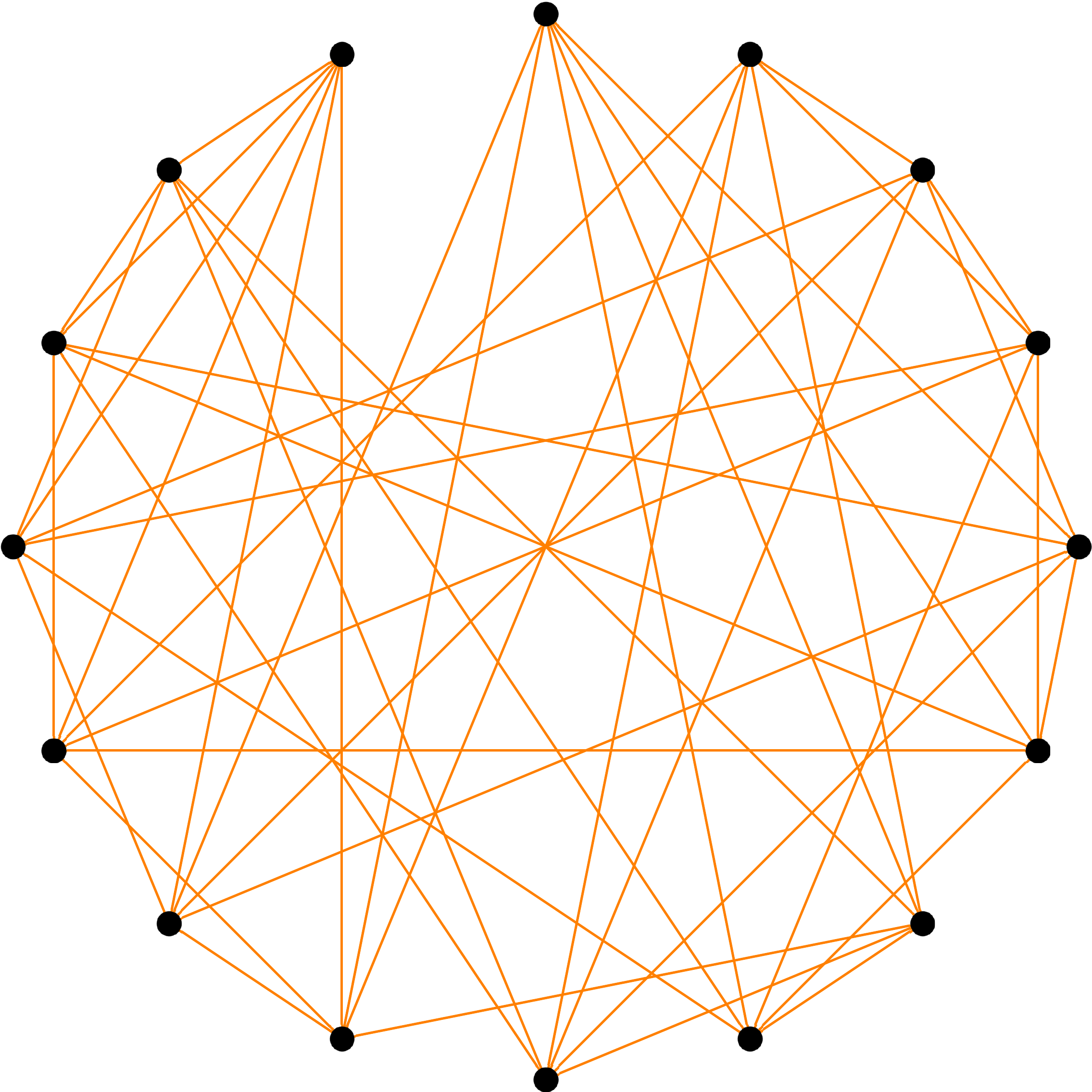}}
	\caption{Two 16-node SRGs with parameters (16,6,2,2).}
	\label{Fig: SRG16A_B}
\end{figure}



\begin{table}[H]
	\centering
	\caption{Similarity scores for comparisons  between different SRGs using two reference nodes. The parameters of the SRGs are as follows: 6a: $SRG$(6,3,0,3), 6b: $SRG$(6,4,2,4), 16a and 16b: $SRG$(16,6,2,2), 45a and 45b: $SRG$(45,12,3,3), 64a and 64b: $SRG$(64,18,2,6).}
	\label{Tab3: Iso2Diff}
	\begin{tabular}{| c | c | c | c | c | }
		\hline
		Graphs & $\epsilon$ Threshold & Euclidean & Euclidean Square & Matusita \\ \hline
		6a vs 6b & 0.3255 & 0.3409 & 0.3522 & 0.3575 \\
		\hline
		16a vs 16b & 0.4828 & 0.8398 & 0.9604 & 0.8191  \\
		\hline
        29a vs 29b & 0.9939 & 0.9997 & 0.99995 & 0.9999\\
        \hline
        36a vs 36b & 0.9357 & 0.9994 & 0.9987 & 0.9994 \\
        \hline
		45a vs 45b & 0.9722 & 0.9962 & 0.9965 & 0.9945 \\
		\hline
		64a vs 64b & 0.8800 & 0.9996 & 0.9986 & 0.9996 \\
        \hline
	\end{tabular}
\end{table}

\begin{figure}[t]
	\centering
    \subfloat[]{\includegraphics[width=0.45\textwidth]{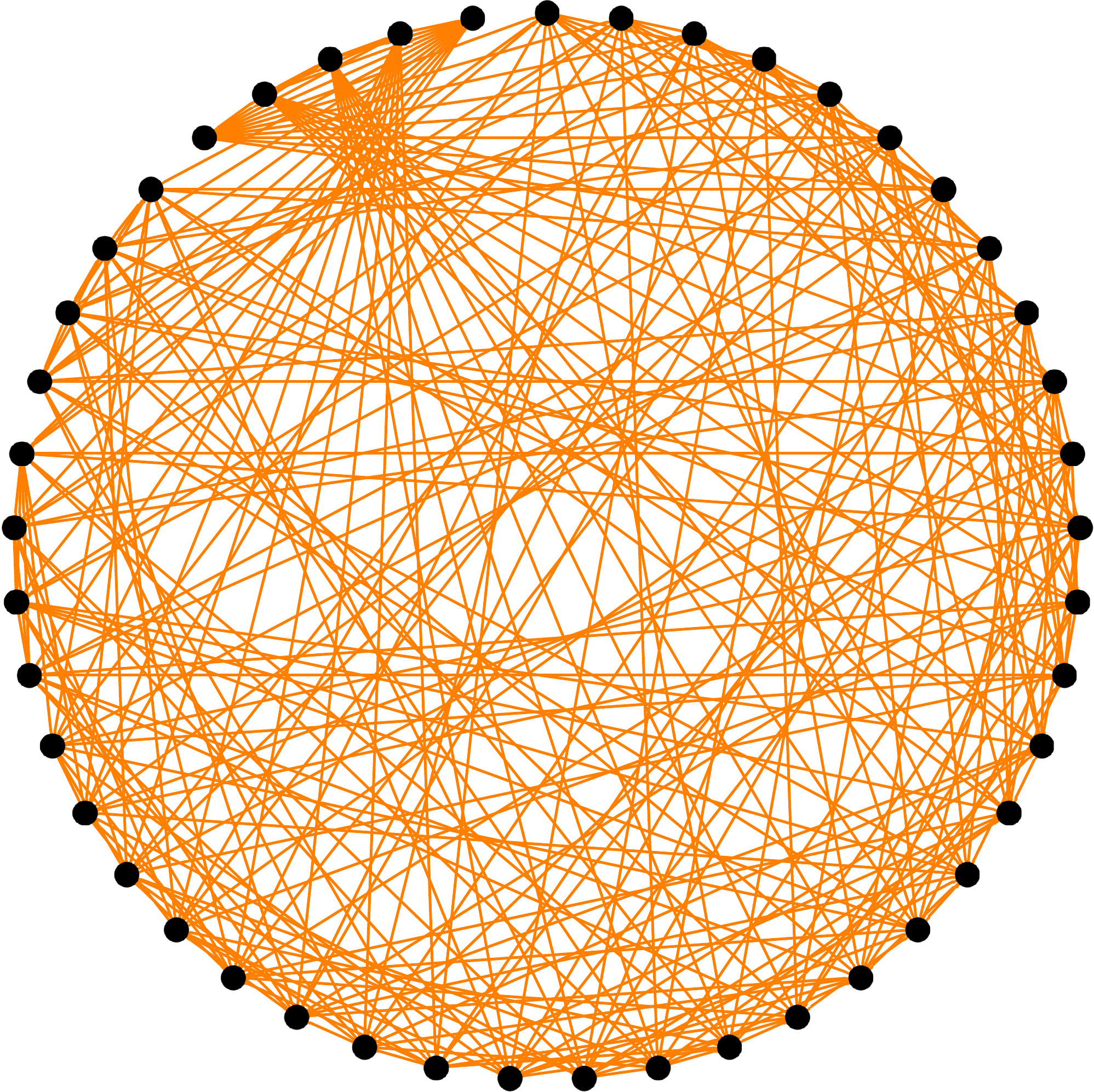}}\quad
    \subfloat[]{\includegraphics[width=0.45\textwidth]{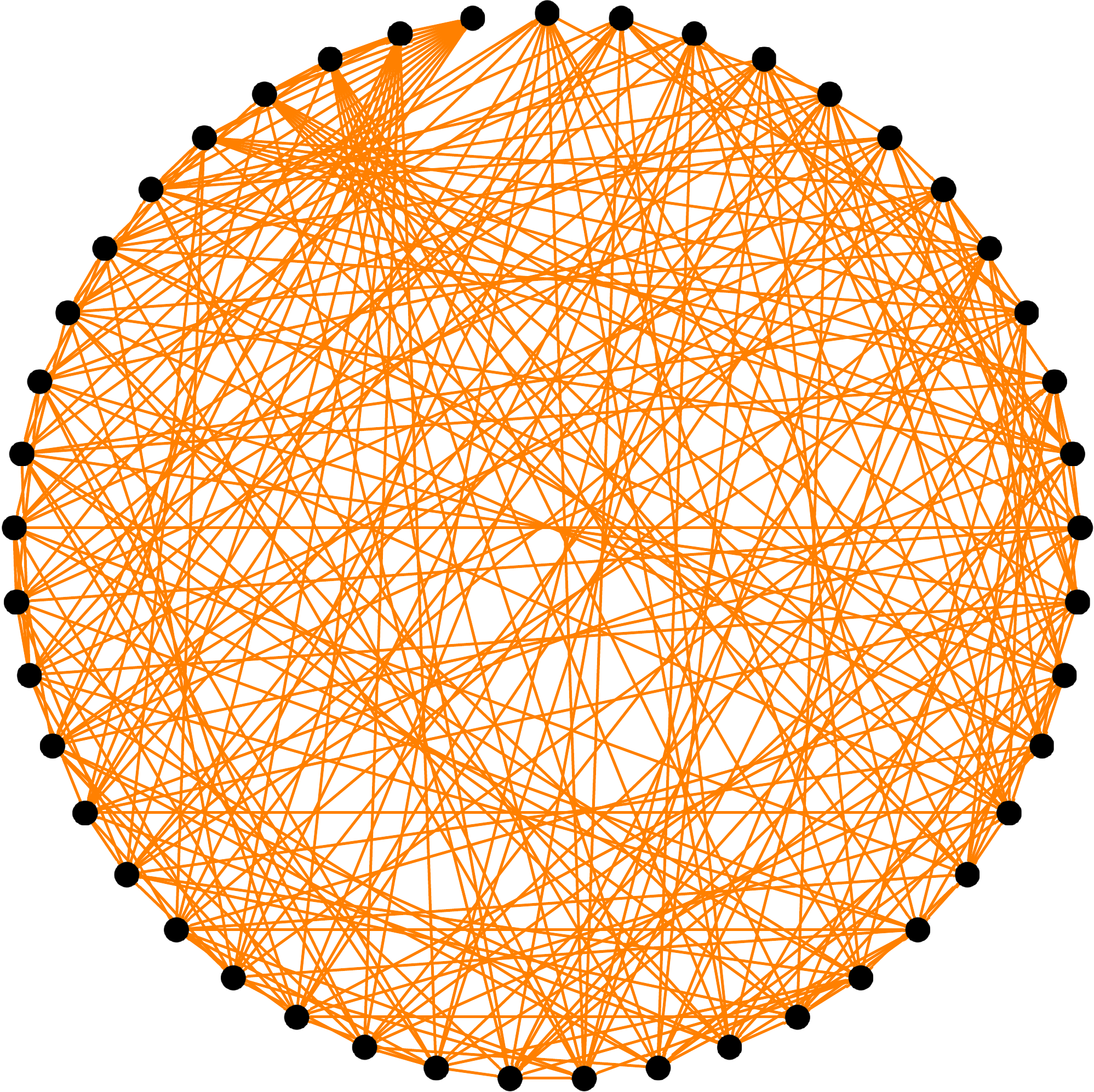}}
	\caption{Two 45 node SRGs with parameters (45,12,3,3).}
	\label{Fig: SRG45A_B}
\end{figure}

Table \ref{Tab3: Iso2Diff} gives the resulting similarity score when comparing a number of different SRGs from the same family. These same graphs have also been compared to themselves, all returning a similarity of 1. Thus, the algorithm successfully reproduces the identity axiom for these SRGs. Note that it is also expected that as the number of nodes increases, the similarity score will increase.



\subsection{Graph Similarity Testing}
\label{Sec: Similarity}

To test the graph similarity metric, we make comparisons between an input graph and a modified copy of this graph. Changes are made by randomly removing edges from the original graph. The graph similarity score is expected to decrease as edges are removed.  Figure \ref{Fig: RemAll} demonstrate this trend for a variety of input graphs. For each number of edges removed, a statistical ensemble of graphs were generated for comparison.  The variation in each of these sets of comparison scores would be due to the variations in structure created by removing different edges. The comparison score for each set is averaged, given by the solid lines in the plots. As we are more interested in how the similarity metric responds to small changes in graph structure, for each data set, we use a smaller step size at the beginning when removing edges. 

As each input graph is generated with respect to the original graph, we can match the nodes between the graphs. This allows us to compare our algorithm with a classical graph similarity technique, such as \textsc{DeltaCon}. Note that we did not conduct this comparison with \textsc{DeltaCon} for SRGs in the previous section, due to the exponential time required by \textsc{DeltaCon} to match the nodes between the comparison graphs.

\begin{figure}[p]
	\centering
    \subfloat[Similarity comparison while removing edges in $ER_A(50, 0.1)$, with 118 edges.\label{Fig: RemER50A}]{
        \begin{minipage}{0.6\textwidth}
            \includegraphics[width=\textwidth]{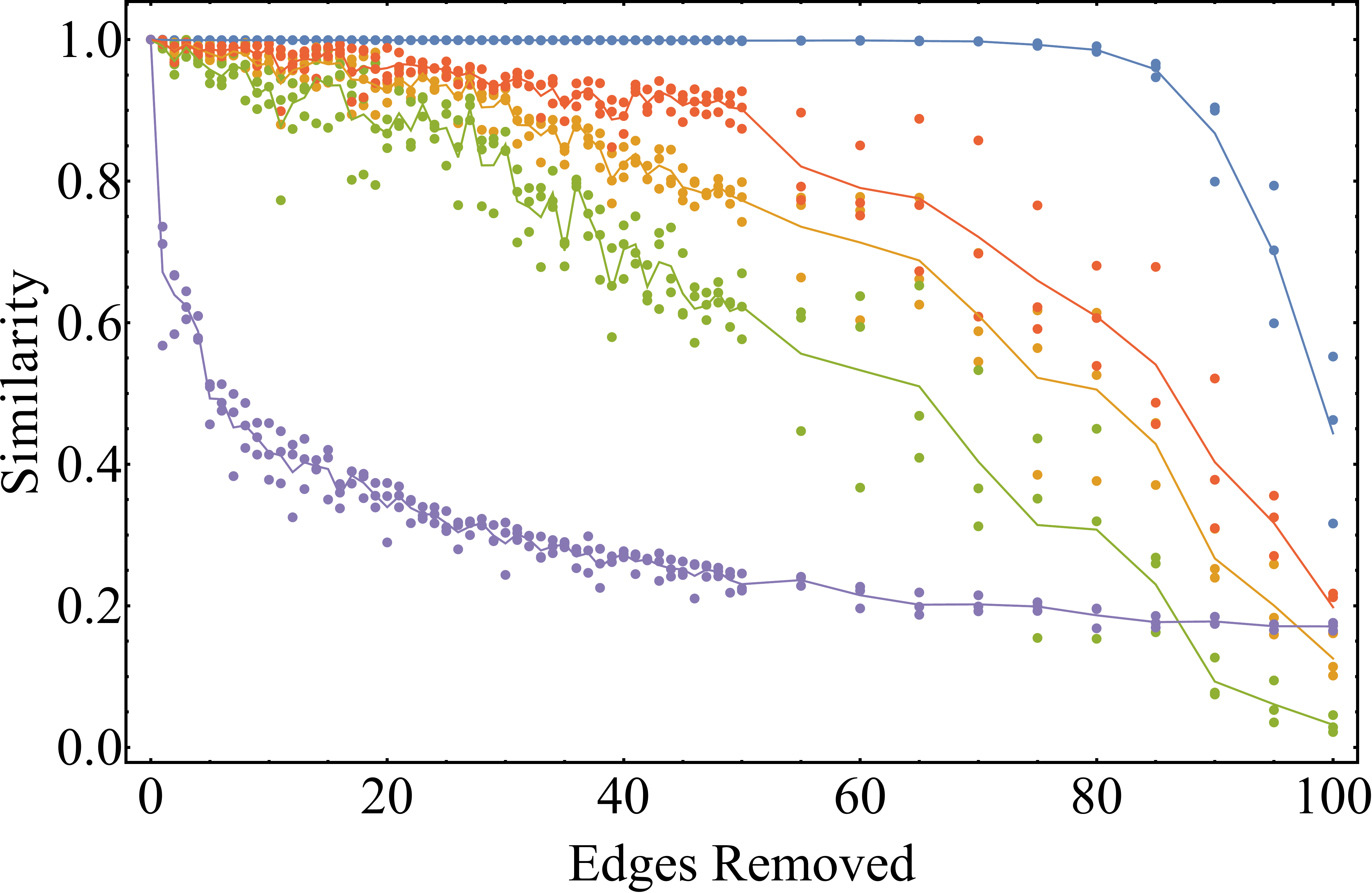}
        \end{minipage}\hfill
        \begin{minipage}{0.36\textwidth}
            \vspace*{\fill}
            \includegraphics[width=\textwidth]{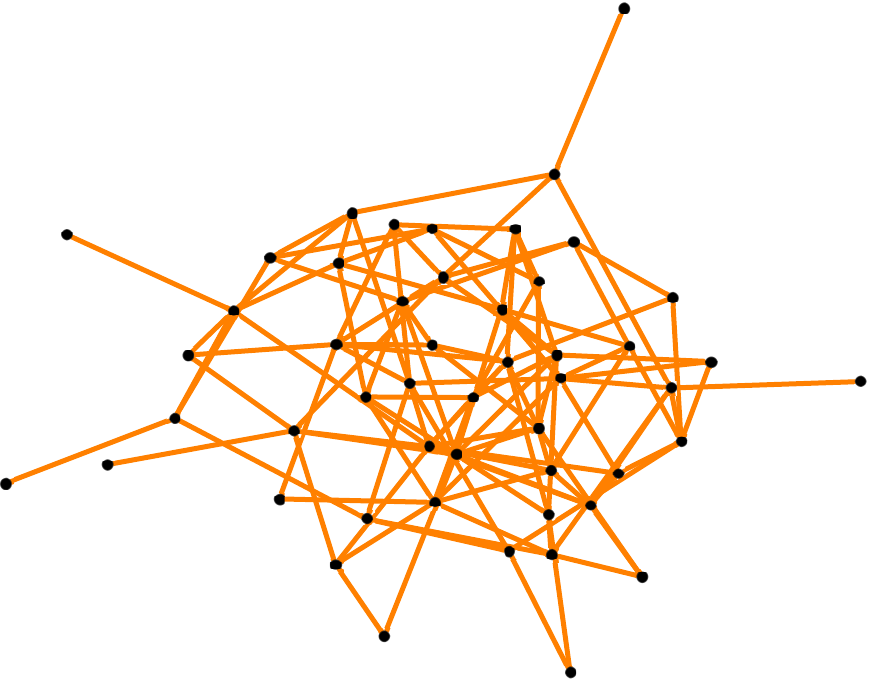}
            \vspace*{\fill}
    	\end{minipage}
    }\\
    \subfloat[Similarity comparison while removing edges in $ER_B(50, 0.1)$, with 118 edges.\label{Fig: RemER50B}]{
        \begin{minipage}{0.6\textwidth}
            \includegraphics[width=\textwidth]{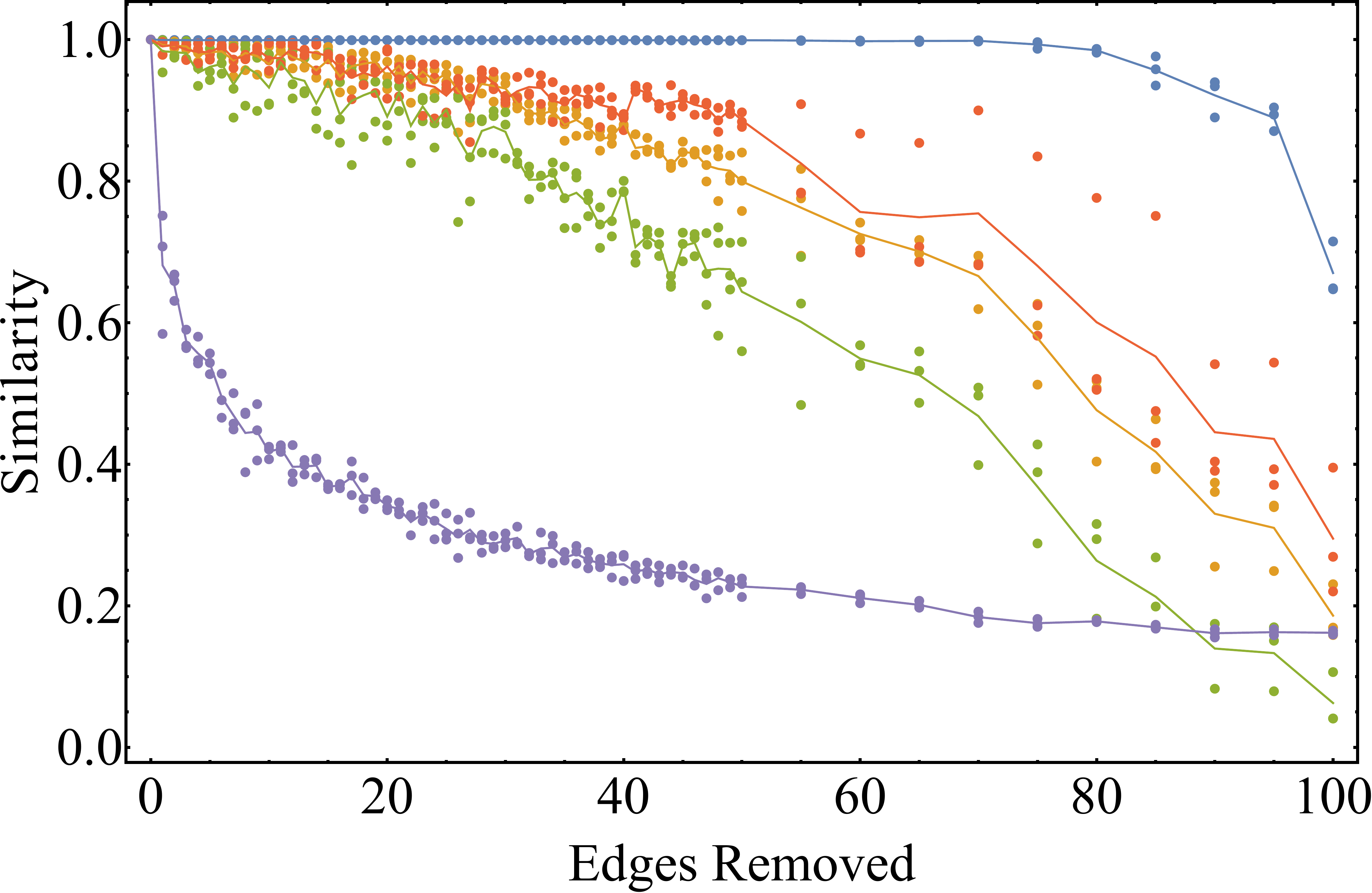}
        \end{minipage}\hfill
        \begin{minipage}{0.36\textwidth}
            \vspace*{\fill}
            \includegraphics[width=\textwidth]{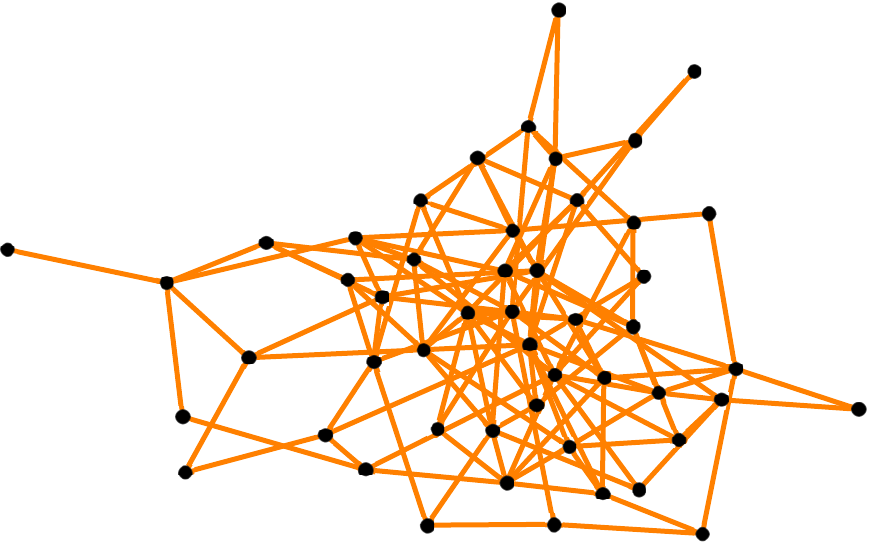}
            \vspace*{\fill}
    	\end{minipage}
    }\\
    \subfloat[Similarity comparison while removing edges in $SF_A(30, 3)$, with 84 edges.\label{Fig: Rem30SFA3}]{
        \begin{minipage}{0.58\textwidth}
            \includegraphics[width=\textwidth]{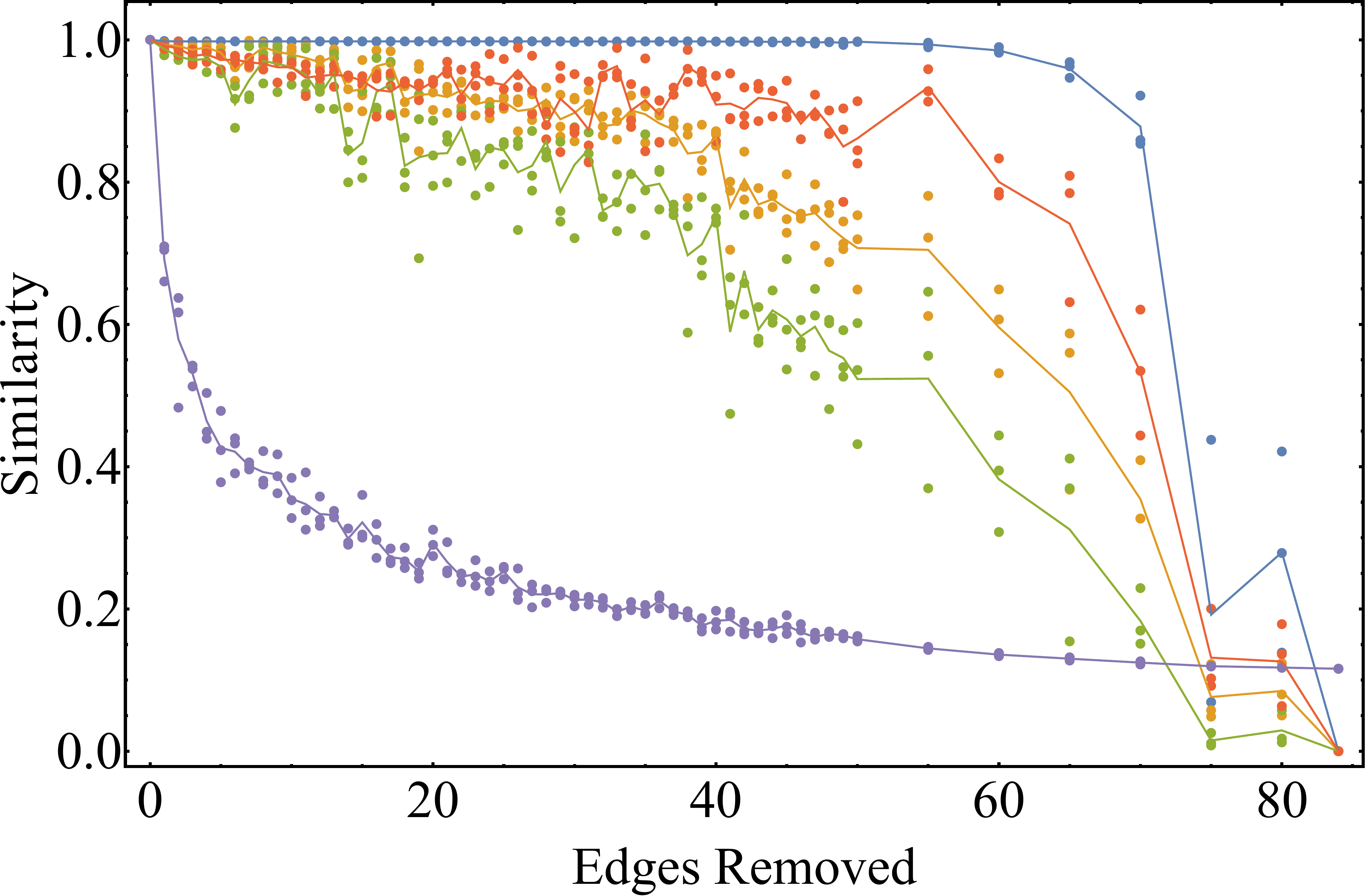}
        \end{minipage}\hfill
        \begin{minipage}{0.34\textwidth}
            \vspace*{\fill}
            \includegraphics[width=\textwidth]{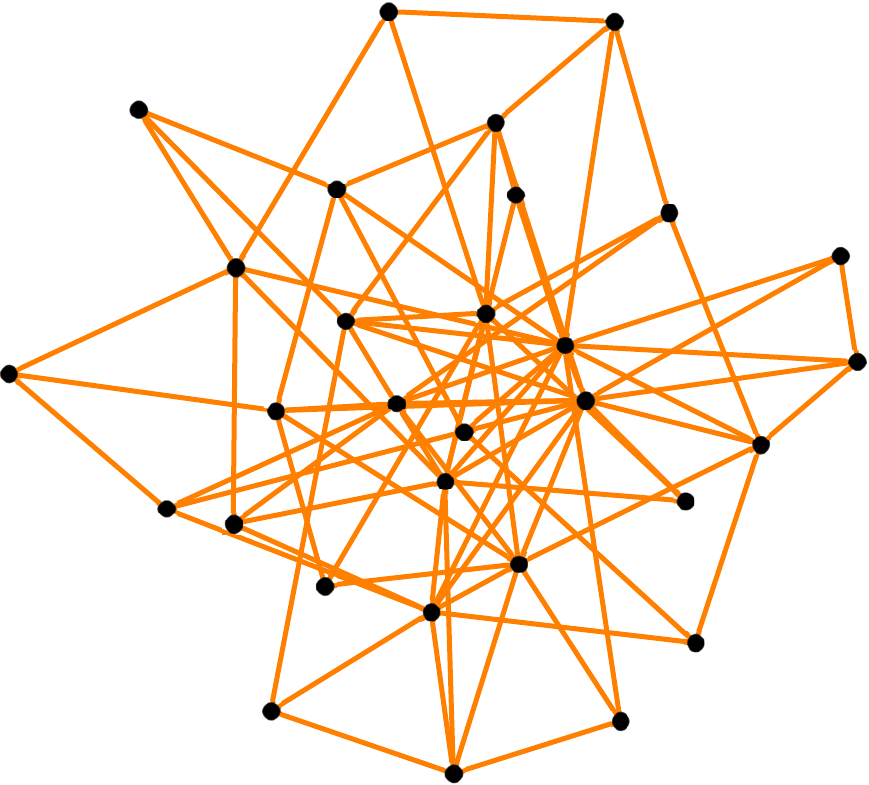}
            \vspace*{\fill}
    	\end{minipage}
    }
    \caption{Resulting similarity score after removing edges from a graph (shown to the right). Results are shown for $\epsilon$ thresholding (blue), Euclidean distance (orange), Euclidean distance squared (green), Matusita distance (red) and \textsc{DeltaCon} (purple). Solid lines denote average score.}
\end{figure}
\begin{figure}[p]
	\ContinuedFloat
	\centering
    \subfloat[Similarity comparison while removing edges in $SF_B(30, 3)$, with 84 edges.\label{Fig: Rem30SFB3}]{
        \begin{minipage}{0.58\textwidth}
            \includegraphics[width=\textwidth]{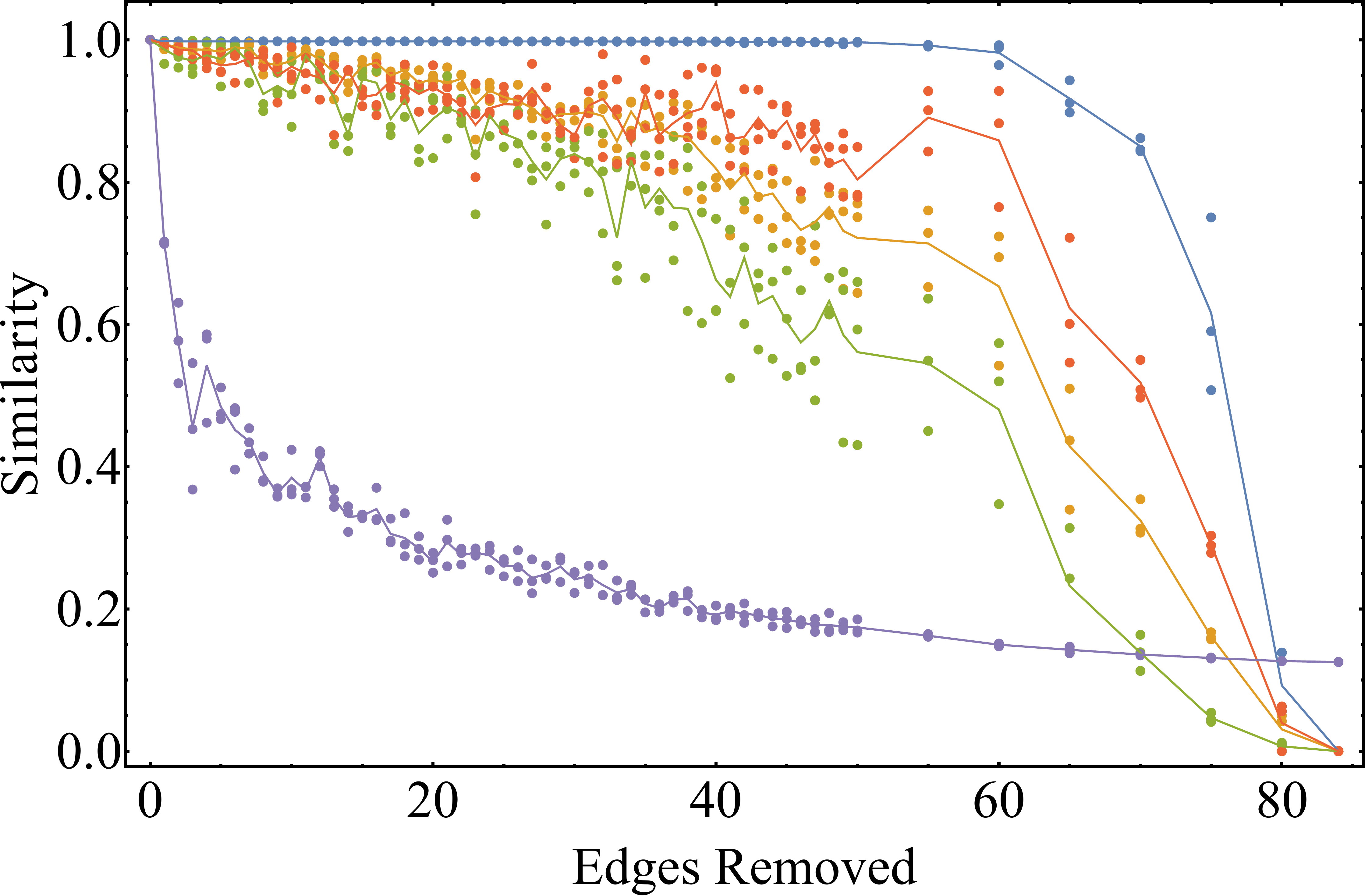}
        \end{minipage}\hfill
        \begin{minipage}{0.34\textwidth}
            \vspace*{\fill}
            \includegraphics[width=\textwidth]{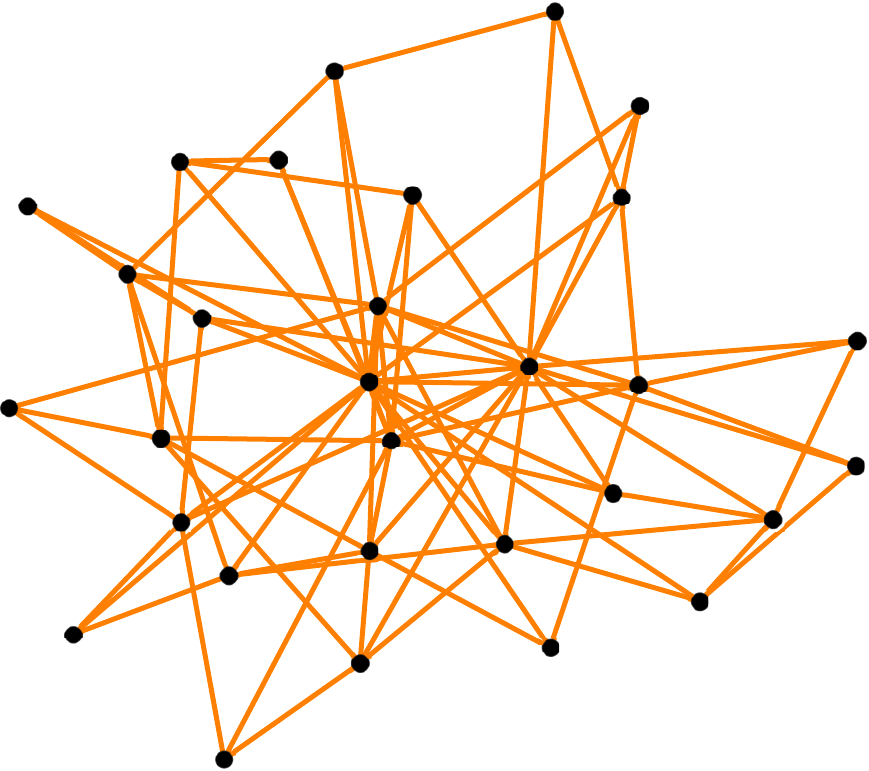}
            \vspace*{\fill}
    	\end{minipage}
    }\\
    \subfloat[Similarity comparison while removing edges in $SF_A(50, 2)$, with 97 edges.\label{Fig: Rem50SFA2}]{
        \begin{minipage}{0.58\textwidth}
            \includegraphics[width=\textwidth]{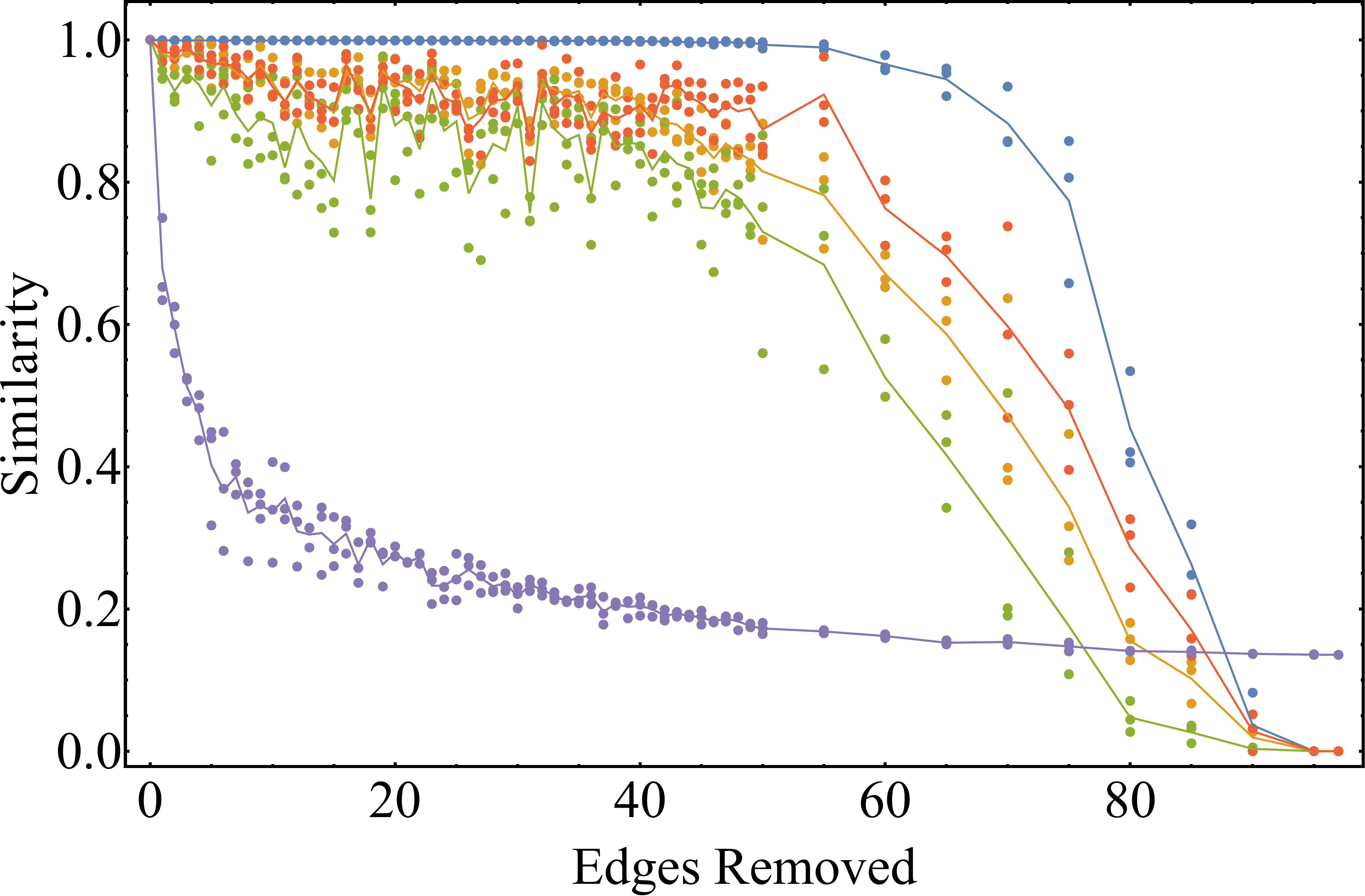}
        \end{minipage}\hfill
        \begin{minipage}{0.34\textwidth}
            \vspace*{\fill}
            \includegraphics[width=\textwidth]{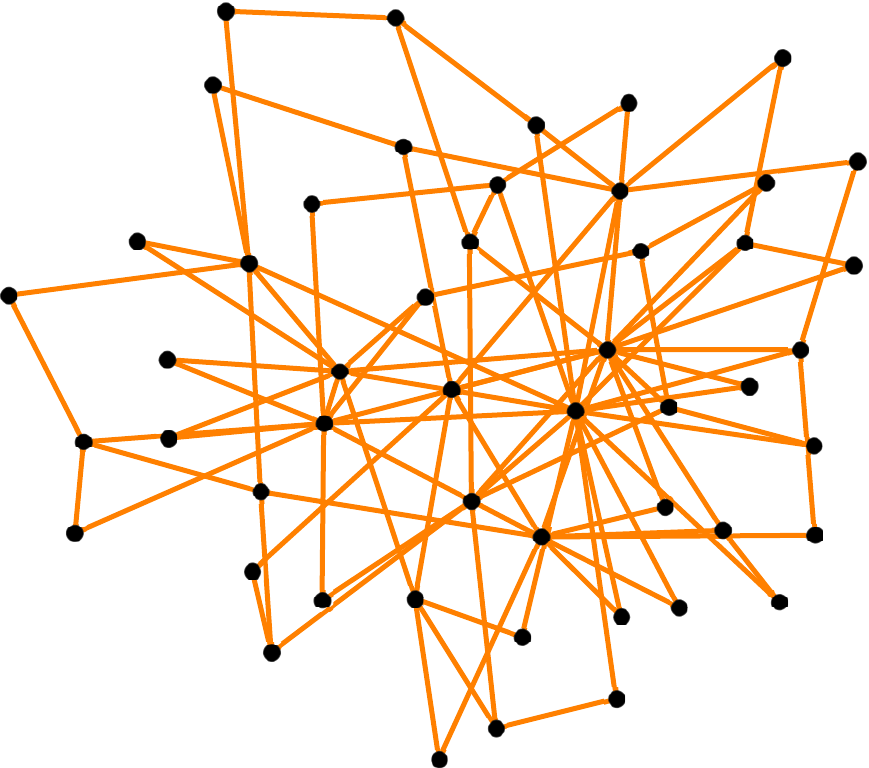}
            \vspace*{\fill}
    	\end{minipage}
    }\\
        \subfloat[Similarity comparison while removing edges in $SF_B(50, 2)$, with 97 edges.\label{Fig: Rem50SFB2}]{
        \begin{minipage}{0.58\textwidth}
            \includegraphics[width=\textwidth]{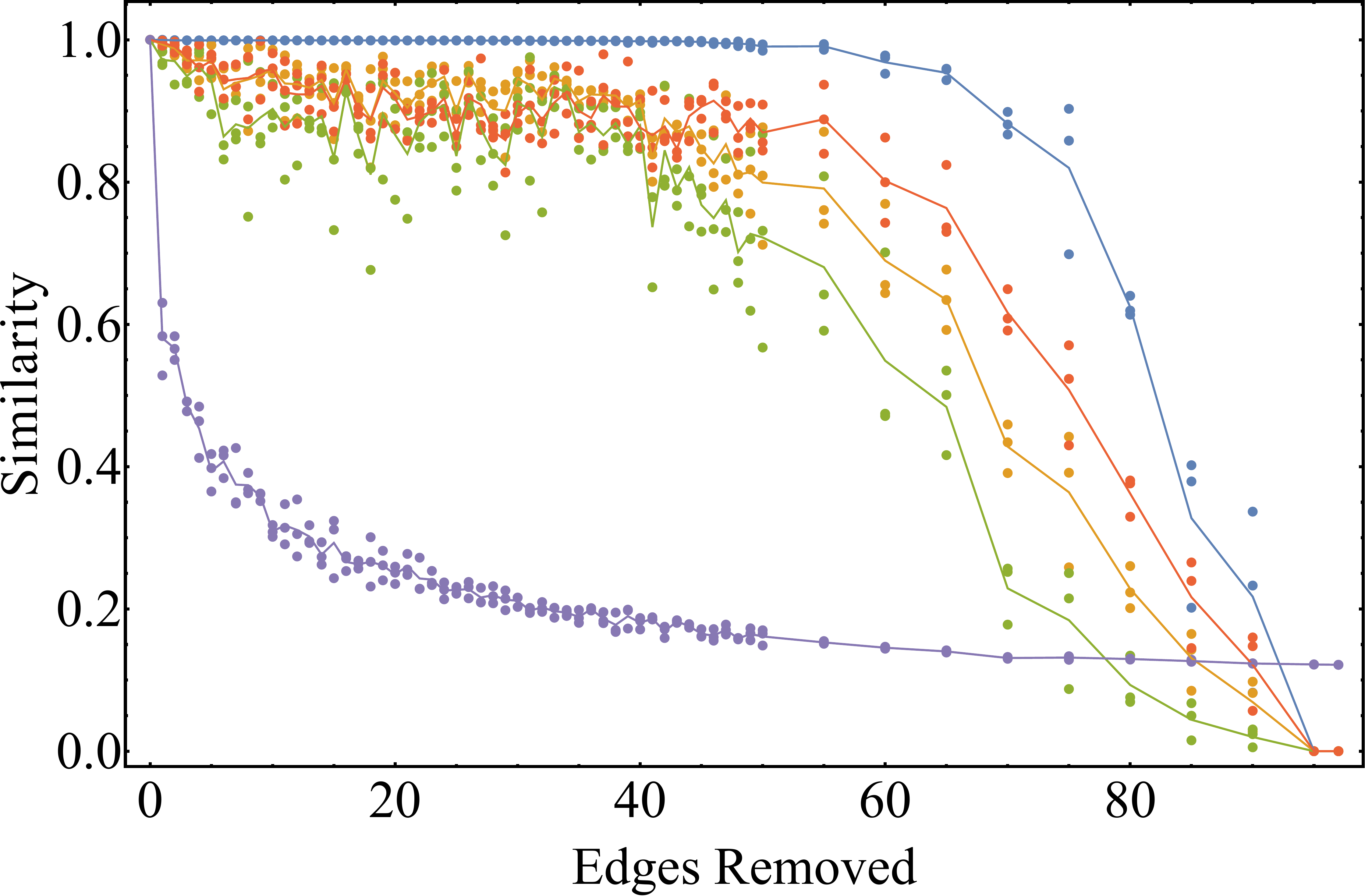}
        \end{minipage}\hfill
        \begin{minipage}{0.34\textwidth}
            \vspace*{\fill}
            \includegraphics[width=\textwidth]{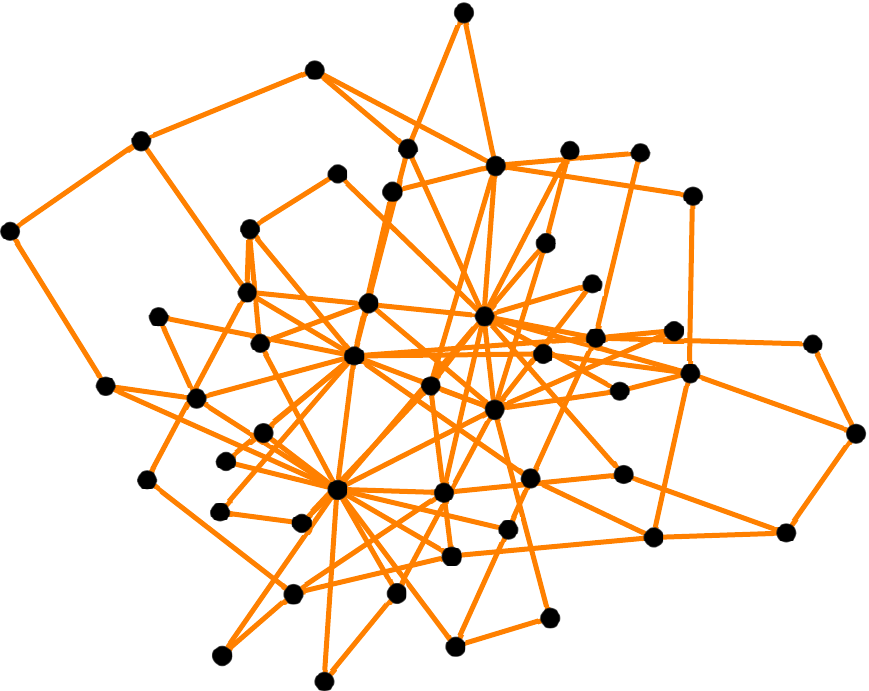}
            \vspace*{\fill}
    	\end{minipage}
    }
	\caption{Resulting similarity score after removing edges from a graph (shown to the right). Results are shown for $\epsilon$ thresholding (blue), Euclidean distance (orange), Euclidean distance squared (green), Matusita distance (red) and \textsc{DeltaCon} (purple). Solid lines denote average score (cont.).}
	\label{Fig: RemAll}
\end{figure}

\section{Discussion and Conclusion}\label{Sec:conclude}

%

It is clear from Figure~\ref{Fig: RemAll} that thresholding the distances does not provide sufficient sensitivity in the comparison measure. The comparison score remains at approximately 0.9 until a sudden drop to 0. As such, this distance metric would likely only be suitable for distinguishing between significantly different graphs.  The other distance measures all give a gradual decrease in similarity score, decreasing faster for the  larger number of changes (denoted as $\nChanges$). This is in contrast to the \textsc{DeltaCon} measure, which is more sensitive for small $\nChanges$ and begins to plateau as $\nChanges$ increases. These two profiles likely represent different approaches to the graph similarity concept. The coined quantum walk metric gives results which are possibly more intuitive, as the graph structure is not likely  to change significantly for small $\nChanges$ (at least not for \ErdosRenyi graphs) so the similarity should gradually decrease until there are no more edges to be deleted. However, \textsc{DeltaCon} suggests that making a few changes has a greater influence on the graph structure with respect to the previous set than for large $\nChanges$. Whether this model is better than our quantum walk inspired model would depend on how the similarity between graphs is defined and the application requirements.

There is also a clear divide between Matusita and Euclidean distance measures, with the Matusita distance giving the least sensitive response. This metric boosts the difference in probabilities, which would increase the comparison score between graphs. However, this does not translate to a greater sensitivity in the similarity comparison, likely due to the boosted score being counteracted by the normalisation process.
The Euclidean measures give roughly the same response as the Matusita measure for small $\nChanges$, but becomes more sensitive as $\nChanges$ increases. The Euclidean square measure has greater sensitivity, but at the cost of having a greater variation. This suggests that this measure is either more sensitive to graph structure changes or less accurate.  Due to the fact that \textsc{DeltaCon} is relatively consistent, it is more likely the latter. The square metric could be useful if the application requires sensitivity to small change in graph structure only, otherwise the Euclidean measure is likely more suitable.

Overall, the coined quantum walk algorithm does not provide very good sensitivity for small $\nChanges$; however, this could be attributed to these changes resulting in little difference in overall graph structure leading to a very similar score. In some cases, such as Figure \ref{Fig: Rem50SFA2}, the similarity score is relatively constant for a large range of $\nChanges$ (in this case, 10 to 50 changes). 
The algorithm however is able to distinguish very similar graphs from less similar graphs. This would be useful in applications which do not require a high sensitivity for small $\nChanges$, such as in machine learning or pattern recognition tasks. It could also be used to aid other processes, such as a preprocessing step to filter out comparisons which are significantly different, while another algorithm (which may have a greater time complexity) is used for a finer sensitivity. 


The coined quantum walk algorithm successfully satisfies the three axioms described by Koutra et al. The identity algorithm has been shown to work for all investigated distance measures using the two reference node scheme. The second axiom is also fulfilled due to the algorithm design, as the input graph order has no effect on the final result. For two graphs with decreasing similarity, the graph comparison score will increase as a reduced number of reference node combinations provide a similar response. As two graphs tend towards being completely different (i.e. a complete graph compared with an empty graph), the comparison score will tend towards $O(n^4)$. As this comparison score is likely much greater than the normalisation between each individual graph, the similarity between the graphs would become negligible. This has been confirmed by comparing complete and empty graphs, which return similarity scores of order $O(10^{-18})$, which is within precision error. The axiom is also substantiated by the results in Figure \ref{Fig: RemAll}, as the similarity score tends to zero as $\nChanges$ increases. This can be compared to the classical \textsc{DeltaCon} algorithm, which plateaus at some similarity score.

In conclusion, this paper proposes a quantum walk inspired algorithm designed to compute graph similarity measures in polynomial time with complexity $O(n^9)$. The algorithm is capable of determining minor changes to significantly similar graph structure, such as determining isomorphism in strongly regular graphs. The algorithm also satisfies the axioms outlined by Koutra et al. \cite{DeltaCon}. 
It is worth noting though that the proposed quantum walk model only works for undirected graphs. Consequently, the proposed algorithm to compute graph similarity measure naturally has this restriction as well. Alternative quantum walk models would need to be explored to develop an algorithm that is compatible for directed graphs.


\section{Acknowledgements}
This work was supported by resources provided by The Pawsey Supercomputing Centre with funding from the Australian Government and the Government of Western Australia. YL acknowledges funding from the National Sciences and Engineering Research Council of Canada.


\begin{thebibliography}{27}%
\makeatletter
\providecommand \@ifxundefined [1]{%
 \@ifx{#1\undefined}
}%
\providecommand \@ifnum [1]{%
 \ifnum #1\expandafter \@firstoftwo
 \else \expandafter \@secondoftwo
 \fi
}%
\providecommand \@ifx [1]{%
 \ifx #1\expandafter \@firstoftwo
 \else \expandafter \@secondoftwo
 \fi
}%
\providecommand \natexlab [1]{#1}%
\providecommand \enquote  [1]{``#1''}%
\providecommand \bibnamefont  [1]{#1}%
\providecommand \bibfnamefont [1]{#1}%
\providecommand \citenamefont [1]{#1}%
\providecommand \href@noop [0]{\@secondoftwo}%
\providecommand \href [0]{\begingroup \@sanitize@url \@href}%
\providecommand \@href[1]{\@@startlink{#1}\@@href}%
\providecommand \@@href[1]{\endgroup#1\@@endlink}%
\providecommand \@sanitize@url [0]{\catcode `\\12\catcode `\$12\catcode
  `\&12\catcode `\#12\catcode `\^12\catcode `\_12\catcode `\%12\relax}%
\providecommand \@@startlink[1]{}%
\providecommand \@@endlink[0]{}%
\providecommand \url  [0]{\begingroup\@sanitize@url \@url }%
\providecommand \@url [1]{\endgroup\@href {#1}{\urlprefix }}%
\providecommand \urlprefix  [0]{URL }%
\providecommand \Eprint [0]{\href }%
\providecommand \doibase [0]{http://dx.doi.org/}%
\providecommand \selectlanguage [0]{\@gobble}%
\providecommand \bibinfo  [0]{\@secondoftwo}%
\providecommand \bibfield  [0]{\@secondoftwo}%
\providecommand \translation [1]{[#1]}%
\providecommand \BibitemOpen [0]{}%
\providecommand \bibitemStop [0]{}%
\providecommand \bibitemNoStop [0]{.\EOS\space}%
\providecommand \EOS [0]{\spacefactor3000\relax}%
\providecommand \BibitemShut  [1]{\csname bibitem#1\endcsname}%
\let\auto@bib@innerbib\@empty
\bibitem [{\citenamefont {Ek}\ \emph {et~al.}(2015)\citenamefont {Ek},
  \citenamefont {VerSchneider}, \citenamefont {Cahill},\ and\ \citenamefont
  {Narayan}}]{socialNetwork}%
  \BibitemOpen
  \bibfield  {author} {\bibinfo {author} {\bibfnamefont {Bryan}\ \bibnamefont
  {Ek}}, \bibinfo {author} {\bibfnamefont {Caitlin}\ \bibnamefont
  {VerSchneider}}, \bibinfo {author} {\bibfnamefont {Nathan}\ \bibnamefont
  {Cahill}}, \ and\ \bibinfo {author} {\bibfnamefont {Darren}\ \bibnamefont
  {Narayan}},\ }\bibfield  {title} {\enquote {\bibinfo {title} {A comprehensive
  comparison of graph theory metrics for social networks},}\ }\href@noop {}
  {\bibfield  {journal} {\bibinfo  {journal} {Social Network Analysis and
  Mining}\ }\textbf {\bibinfo {volume} {5}},\ \bibinfo {pages} {1--7} (\bibinfo
  {year} {2015})}\BibitemShut {NoStop}%
\bibitem [{\citenamefont {Horv\'{a}th}\ \emph {et~al.}(2004)\citenamefont
  {Horv\'{a}th}, \citenamefont {G\"{a}rtner},\ and\ \citenamefont
  {Wrobel}}]{ML}%
  \BibitemOpen
  \bibfield  {author} {\bibinfo {author} {\bibfnamefont {Tam\'{a}s}\
  \bibnamefont {Horv\'{a}th}}, \bibinfo {author} {\bibfnamefont {Thomas}\
  \bibnamefont {G\"{a}rtner}}, \ and\ \bibinfo {author} {\bibfnamefont
  {Stefan}\ \bibnamefont {Wrobel}},\ }\bibfield  {title} {\enquote {\bibinfo
  {title} {Cyclic pattern kernels for predictive graph mining},}\ }in\
  \href@noop {} {\emph {\bibinfo {booktitle} {Proceedings of the Tenth ACM
  SIGKDD International Conference on Knowledge Discovery and Data Mining}}},\
  \bibinfo {series and number} {KDD '04}\ (\bibinfo  {publisher} {ACM},\
  \bibinfo {year} {2004})\ pp.\ \bibinfo {pages} {158--167}\BibitemShut
  {NoStop}%
\bibitem [{\citenamefont {Luo}\ \emph {et~al.}(2010)\citenamefont {Luo},
  \citenamefont {Zhao}, \citenamefont {Cheng}, \citenamefont {Jiang},\ and\
  \citenamefont {Wang}}]{protein-protein-interaction}%
  \BibitemOpen
  \bibfield  {author} {\bibinfo {author} {\bibfnamefont {Yong}\ \bibnamefont
  {Luo}}, \bibinfo {author} {\bibfnamefont {Yan}\ \bibnamefont {Zhao}},
  \bibinfo {author} {\bibfnamefont {Lei}\ \bibnamefont {Cheng}}, \bibinfo
  {author} {\bibfnamefont {Ping}\ \bibnamefont {Jiang}}, \ and\ \bibinfo
  {author} {\bibfnamefont {Jianxin}\ \bibnamefont {Wang}},\ }\bibfield  {title}
  {\enquote {\bibinfo {title} {Protein-protein interaction network comparison
  based on wavelet and principal component analysis},}\ \ }(\bibinfo
  {publisher} {IEEE Publishing},\ \bibinfo {year} {2010})\ pp.\ \bibinfo
  {pages} {294--298}\BibitemShut {NoStop}%
\bibitem [{\citenamefont {Mheich}\ \emph {et~al.}(2017)\citenamefont {Mheich},
  \citenamefont {Hassan}, \citenamefont {Khalil}, \citenamefont {Gripon},
  \citenamefont {Dufor},\ and\ \citenamefont {Wendling}}]{brainNetwork}%
  \BibitemOpen
  \bibfield  {author} {\bibinfo {author} {\bibfnamefont {Ahmad}\ \bibnamefont
  {Mheich}}, \bibinfo {author} {\bibfnamefont {Mahmoud}\ \bibnamefont
  {Hassan}}, \bibinfo {author} {\bibfnamefont {Mohamad}\ \bibnamefont
  {Khalil}}, \bibinfo {author} {\bibfnamefont {Vincent}\ \bibnamefont
  {Gripon}}, \bibinfo {author} {\bibfnamefont {Olivier}\ \bibnamefont {Dufor}},
  \ and\ \bibinfo {author} {\bibfnamefont {Fabrice}\ \bibnamefont {Wendling}},\
  }\bibfield  {title} {\enquote {\bibinfo {title} {Siminet: a novel method for
  quantifying brain network similarity},}\ }\href@noop {} {\bibfield  {journal}
  {\bibinfo  {journal} {IEEE transactions on pattern analysis and machine
  intelligence}\ } (\bibinfo {year} {2017})}\BibitemShut {NoStop}%
\bibitem [{\citenamefont {Zheng}\ \emph {et~al.}(2015)\citenamefont {Zheng},
  \citenamefont {Zou}, \citenamefont {Lian}, \citenamefont {Wang},\ and\
  \citenamefont {Zhao}}]{DatabaseComp}%
  \BibitemOpen
  \bibfield  {author} {\bibinfo {author} {\bibfnamefont {Weiguo}\ \bibnamefont
  {Zheng}}, \bibinfo {author} {\bibfnamefont {Lei}\ \bibnamefont {Zou}},
  \bibinfo {author} {\bibfnamefont {Xiang}\ \bibnamefont {Lian}}, \bibinfo
  {author} {\bibfnamefont {Dong}\ \bibnamefont {Wang}}, \ and\ \bibinfo
  {author} {\bibfnamefont {Dongyan}\ \bibnamefont {Zhao}},\ }\bibfield  {title}
  {\enquote {\bibinfo {title} {Efficient graph similarity search over large
  graph databases},}\ }\href@noop {} {\ \textbf {\bibinfo {volume} {27}},\
  \bibinfo {pages} {964--978} (\bibinfo {year} {2015})}\BibitemShut {NoStop}%
\bibitem [{\citenamefont {Showbridge}\ \emph {et~al.}(1999)\citenamefont
  {Showbridge}, \citenamefont {Kraetzl},\ and\ \citenamefont
  {Ray}}]{DetectionNetwork}%
  \BibitemOpen
  \bibfield  {author} {\bibinfo {author} {\bibfnamefont {P.}~\bibnamefont
  {Showbridge}}, \bibinfo {author} {\bibfnamefont {M.}~\bibnamefont {Kraetzl}},
  \ and\ \bibinfo {author} {\bibfnamefont {D.}~\bibnamefont {Ray}},\ }\bibfield
   {title} {\enquote {\bibinfo {title} {Detection of abnormal change in dynamic
  networks},}\ \ }(\bibinfo  {publisher} {IEEE Publishing},\ \bibinfo {year}
  {1999})\ pp.\ \bibinfo {pages} {557--562}\BibitemShut {NoStop}%
\bibitem [{\citenamefont {{Koutra}}\ \emph {et~al.}(2013)\citenamefont
  {{Koutra}}, \citenamefont {{Vogelstein}},\ and\ \citenamefont
  {{Faloutsos}}}]{DeltaCon}%
  \BibitemOpen
  \bibfield  {author} {\bibinfo {author} {\bibfnamefont {D.}~\bibnamefont
  {{Koutra}}}, \bibinfo {author} {\bibfnamefont {J.~T.}\ \bibnamefont
  {{Vogelstein}}}, \ and\ \bibinfo {author} {\bibfnamefont {C.}~\bibnamefont
  {{Faloutsos}}},\ }\bibfield  {title} {\enquote {\bibinfo {title} {{DELTACON:
  A principled massive-graph similarity function}},}\ }\href@noop {} {\bibfield
   {journal} {\bibinfo  {journal} {arXiv:1304.4657}\ } (\bibinfo {year}
  {2013})}\BibitemShut {NoStop}%
\bibitem [{\citenamefont {Bunke}(2000)}]{Bunkegraphmatching}%
  \BibitemOpen
  \bibfield  {author} {\bibinfo {author} {\bibfnamefont {Horst}\ \bibnamefont
  {Bunke}},\ }\bibfield  {title} {\enquote {\bibinfo {title} {Graph matching:
  Theoretical foundations, algorithms, and applications},}\ }in\ \href@noop {}
  {\emph {\bibinfo {booktitle} {In Proceedings of Vision Interface 2000,
  Montreal}}}\ (\bibinfo {year} {2000})\ pp.\ \bibinfo {pages}
  {82--88}\BibitemShut {NoStop}%
\bibitem [{\citenamefont {Emmert-Streib}\ \emph {et~al.}(2016)\citenamefont
  {Emmert-Streib}, \citenamefont {Dehmer},\ and\ \citenamefont
  {Shi}}]{50years}%
  \BibitemOpen
  \bibfield  {author} {\bibinfo {author} {\bibfnamefont {Frank}\ \bibnamefont
  {Emmert-Streib}}, \bibinfo {author} {\bibfnamefont {Matthias}\ \bibnamefont
  {Dehmer}}, \ and\ \bibinfo {author} {\bibfnamefont {Yongtang}\ \bibnamefont
  {Shi}},\ }\bibfield  {title} {\enquote {\bibinfo {title} {Fifty years of
  graph matching, network alignment and network comparison},}\ }\href@noop {}
  {\bibfield  {journal} {\bibinfo  {journal} {Information Sciences}\ }\textbf
  {\bibinfo {volume} {346-347}},\ \bibinfo {pages} {180--197} (\bibinfo {year}
  {2016})}\BibitemShut {NoStop}%
\bibitem [{\citenamefont {Levi}(1973)}]{LeviMCS}%
  \BibitemOpen
  \bibfield  {author} {\bibinfo {author} {\bibfnamefont {G.}~\bibnamefont
  {Levi}},\ }\bibfield  {title} {\enquote {\bibinfo {title} {A note on the
  derivation of maximal common subgraphs of two directed or undirected
  graphs},}\ }\href@noop {} {\bibfield  {journal} {\bibinfo  {journal}
  {CALCOLO}\ }\textbf {\bibinfo {volume} {9}},\ \bibinfo {pages} {341--352}
  (\bibinfo {year} {1973})}\BibitemShut {NoStop}%
\bibitem [{\citenamefont {Bunke}\ and\ \citenamefont
  {Shearer}(1998)}]{BunkeMaxComSub}%
  \BibitemOpen
  \bibfield  {author} {\bibinfo {author} {\bibfnamefont {Horst}\ \bibnamefont
  {Bunke}}\ and\ \bibinfo {author} {\bibfnamefont {Kim}\ \bibnamefont
  {Shearer}},\ }\bibfield  {title} {\enquote {\bibinfo {title} {A graph
  distance metric based on the maximal common subgraph},}\ }\href@noop {}
  {\bibfield  {journal} {\bibinfo  {journal} {Pattern Recognition Letters}\
  }\textbf {\bibinfo {volume} {19}},\ \bibinfo {pages} {255--259} (\bibinfo
  {year} {1998})}\BibitemShut {NoStop}%
\bibitem [{\citenamefont {Douglas}\ and\ \citenamefont
  {Wang}(2008)}]{DouglasGI}%
  \BibitemOpen
  \bibfield  {author} {\bibinfo {author} {\bibfnamefont {Brendan~L}\
  \bibnamefont {Douglas}}\ and\ \bibinfo {author} {\bibfnamefont {Jingbo~B}\
  \bibnamefont {Wang}},\ }\bibfield  {title} {\enquote {\bibinfo {title} {A
  classical approach to the graph isomorphism problem using quantum walks},}\
  }\href@noop {} {\bibfield  {journal} {\bibinfo  {journal} {Journal of Physics
  A: Mathematical and Theoretical}\ }\textbf {\bibinfo {volume} {41}} (\bibinfo
  {year} {2008})}\BibitemShut {NoStop}%
\bibitem [{\citenamefont {Douglas}()}]{DouglasPhD}%
  \BibitemOpen
  \bibfield  {author} {\bibinfo {author} {\bibfnamefont {Brendan~Lindsay}\
  \bibnamefont {Douglas}},\ }\href@noop {} {\enquote {\bibinfo {title} {Quantum
  and classical algorithms for graph classification and search problems},}\
  }\BibitemShut {NoStop}%
\bibitem [{\citenamefont {Umeyama}(1988)}]{Umeyama_Eigen}%
  \BibitemOpen
  \bibfield  {author} {\bibinfo {author} {\bibfnamefont {S.}~\bibnamefont
  {Umeyama}},\ }\bibfield  {title} {\enquote {\bibinfo {title} {An
  eigendecomposition approach to weighted graph matching problems},}\
  }\href@noop {} {\bibfield  {journal} {\bibinfo  {journal} {Pattern Analysis
  and Machine Intelligence, IEEE Transactions on}\ }\textbf {\bibinfo {volume}
  {10}},\ \bibinfo {pages} {695--703} (\bibinfo {year} {1988})}\BibitemShut
  {NoStop}%
\bibitem [{\citenamefont {Manouchehri}\ and\ \citenamefont
  {Wang}(2013)}]{JWQuantumWalks}%
  \BibitemOpen
  \bibfield  {author} {\bibinfo {author} {\bibfnamefont {Kia}\ \bibnamefont
  {Manouchehri}}\ and\ \bibinfo {author} {\bibfnamefont {Jingbo}\ \bibnamefont
  {Wang}},\ }\href@noop {} {\emph {\bibinfo {title} {Physical Implementation of
  Quantum Walks}}},\ \bibinfo {edition} {1st}\ ed.,\ Quantum Science and
  Technology\ (\bibinfo  {publisher} {Springer},\ \bibinfo {address} {Berlin,
  Heidelberg},\ \bibinfo {year} {2013})\BibitemShut {NoStop}%
\bibitem [{\citenamefont {Watrous}(2001)}]{Watrous_QW}%
  \BibitemOpen
  \bibfield  {author} {\bibinfo {author} {\bibfnamefont {John}\ \bibnamefont
  {Watrous}},\ }\bibfield  {title} {\enquote {\bibinfo {title} {Quantum
  simulations of classical random walks and undirected graph connectivity},}\
  }\href@noop {} {\bibfield  {journal} {\bibinfo  {journal} {Journal of
  Computer and System Sciences}\ }\textbf {\bibinfo {volume} {62}},\ \bibinfo
  {pages} {376,--391} (\bibinfo {year} {2001})}\BibitemShut {NoStop}%
\bibitem [{\citenamefont {\Erdos}\ and\ \citenamefont
  {\Renyi}(1959)}]{ER_RandomGraphs}%
  \BibitemOpen
  \bibfield  {author} {\bibinfo {author} {\bibfnamefont {P.}~\bibnamefont
  {\Erdos}}\ and\ \bibinfo {author} {\bibfnamefont {A.}~\bibnamefont
  {\Renyi}},\ }\bibfield  {title} {\enquote {\bibinfo {title} {On random
  graphs},}\ }in\ \href@noop {} {\emph {\bibinfo {booktitle} {Publicationes
  Mathematicae}}},\ Vol.~\bibinfo {volume} {6}\ (\bibinfo {year} {1959})\ pp.\
  \bibinfo {pages} {290--297}\BibitemShut {NoStop}%
\bibitem [{\citenamefont {\Erdos}\ and\ \citenamefont
  {\Renyi}(1960)}]{ER_Evolution}%
  \BibitemOpen
  \bibfield  {author} {\bibinfo {author} {\bibfnamefont {P.}~\bibnamefont
  {\Erdos}}\ and\ \bibinfo {author} {\bibfnamefont {A.}~\bibnamefont
  {\Renyi}},\ }\bibfield  {title} {\enquote {\bibinfo {title} {On the evolution
  of random graphs},}\ }in\ \href@noop {} {\emph {\bibinfo {booktitle}
  {Publication of the Mathematical Institute of the Hungarian Academy of
  Sciences}}}\ (\bibinfo {year} {1960})\ pp.\ \bibinfo {pages}
  {17--61}\BibitemShut {NoStop}%
\bibitem [{\citenamefont {Ganesh}\ \emph {et~al.}(2005)\citenamefont {Ganesh},
  \citenamefont {Massoulie},\ and\ \citenamefont {Towsley}}]{ER_Use_Epidemic}%
  \BibitemOpen
  \bibfield  {author} {\bibinfo {author} {\bibfnamefont {A.}~\bibnamefont
  {Ganesh}}, \bibinfo {author} {\bibfnamefont {L.}~\bibnamefont {Massoulie}}, \
  and\ \bibinfo {author} {\bibfnamefont {D.}~\bibnamefont {Towsley}},\
  }\bibfield  {title} {\enquote {\bibinfo {title} {The effect of network
  topology on the spread of epidemics},}\ \ }(\bibinfo  {publisher} {IEEE},\
  \bibinfo {year} {2005})\ pp.\ \bibinfo {pages} {1455--1466}\BibitemShut
  {NoStop}%
\bibitem [{\citenamefont {Barab\'{a}si}\ and\ \citenamefont
  {Albert}(1999)}]{SF_EmergenceScaling}%
  \BibitemOpen
  \bibfield  {author} {\bibinfo {author} {\bibfnamefont
  {Albert-L\'{a}szl\'{o}}\ \bibnamefont {Barab\'{a}si}}\ and\ \bibinfo {author}
  {\bibfnamefont {R\'{e}ka}\ \bibnamefont {Albert}},\ }\bibfield  {title}
  {\enquote {\bibinfo {title} {Emergence of scaling in random networks},}\
  }\href@noop {} {\bibfield  {journal} {\bibinfo  {journal} {Science}\ }\textbf
  {\bibinfo {volume} {286}},\ \bibinfo {pages} {509--512} (\bibinfo {year}
  {1999})}\BibitemShut {NoStop}%
\bibitem [{\citenamefont {Réka~Albert}\ and\ \citenamefont
  {Barabási}(1999)}]{SF_WWW}%
  \BibitemOpen
  \bibfield  {author} {\bibinfo {author} {\bibfnamefont {Hawoong~Jeong}\
  \bibnamefont {Réka~Albert}}\ and\ \bibinfo {author} {\bibfnamefont
  {Albert-László}\ \bibnamefont {Barabási}},\ }\bibfield  {title} {\enquote
  {\bibinfo {title} {Internet: Diameter of the world-wide web},}\ }\href@noop
  {} {\bibfield  {journal} {\bibinfo  {journal} {Nature}\ }\textbf {\bibinfo
  {volume} {401}},\ \bibinfo {pages} {130--131} (\bibinfo {year}
  {1999})}\BibitemShut {NoStop}%
\bibitem [{\citenamefont {Albert}(2005)}]{SF_protein-protein1}%
  \BibitemOpen
  \bibfield  {author} {\bibinfo {author} {\bibfnamefont {R\'{e}ka}\
  \bibnamefont {Albert}},\ }\bibfield  {title} {\enquote {\bibinfo {title}
  {Scale-free networks in cell biology},}\ }\href@noop {} {\bibfield  {journal}
  {\bibinfo  {journal} {Journal of cell science}\ }\textbf {\bibinfo {volume}
  {118}},\ \bibinfo {pages} {4947--4957} (\bibinfo {year} {2005})}\BibitemShut
  {NoStop}%
\bibitem [{\citenamefont {Nafis}\ \emph {et~al.}(2015)\citenamefont {Nafis},
  \citenamefont {Kalaiarasan}, \citenamefont {Brojen~Singh}, \citenamefont
  {Husain},\ and\ \citenamefont {Bamezai}}]{SF_protein-protein2}%
  \BibitemOpen
  \bibfield  {author} {\bibinfo {author} {\bibfnamefont {Shazia}\ \bibnamefont
  {Nafis}}, \bibinfo {author} {\bibfnamefont {Ponnusamy}\ \bibnamefont
  {Kalaiarasan}}, \bibinfo {author} {\bibfnamefont {R.~K.}\ \bibnamefont
  {Brojen~Singh}}, \bibinfo {author} {\bibfnamefont {Mohammad}\ \bibnamefont
  {Husain}}, \ and\ \bibinfo {author} {\bibfnamefont {Rameshwar N.~K.}\
  \bibnamefont {Bamezai}},\ }\bibfield  {title} {\enquote {\bibinfo {title}
  {Apoptosis regulatory protein–protein interaction demonstrates hierarchical
  scale-free fractal network},}\ }\href@noop {} {\bibfield  {journal} {\bibinfo
   {journal} {Briefings in Bioinformatics}\ }\textbf {\bibinfo {volume} {16}},\
  \bibinfo {pages} {675--699} (\bibinfo {year} {2015})}\BibitemShut {NoStop}%
\bibitem [{\citenamefont {He}(2014)}]{SF_BrainActivity}%
  \BibitemOpen
  \bibfield  {author} {\bibinfo {author} {\bibfnamefont {Byj}\ \bibnamefont
  {He}},\ }\bibfield  {title} {\enquote {\bibinfo {title} {Scale-free brain
  activity: past, present, and future},}\ }\href@noop {} {\bibfield  {journal}
  {\bibinfo  {journal} {Trends In Cognitive Sciences}\ }\textbf {\bibinfo
  {volume} {18}},\ \bibinfo {pages} {480--487} (\bibinfo {year}
  {2014})}\BibitemShut {NoStop}%
\bibitem [{\citenamefont {López-Pintado}(2008)}]{SF_Social}%
  \BibitemOpen
  \bibfield  {author} {\bibinfo {author} {\bibfnamefont {Dunia}\ \bibnamefont
  {López-Pintado}},\ }\bibfield  {title} {\enquote {\bibinfo {title}
  {Diffusion in complex social networks},}\ }\href@noop {} {\bibfield
  {journal} {\bibinfo  {journal} {Games and Economic Behavior}\ }\textbf
  {\bibinfo {volume} {62}},\ \bibinfo {pages} {573--590} (\bibinfo {year}
  {2008})}\BibitemShut {NoStop}%
\bibitem [{\citenamefont {Barabási}\ \emph {et~al.}(2000)\citenamefont
  {Barabási}, \citenamefont {Albert},\ and\ \citenamefont
  {Jeong}}]{SF_Characteristics}%
  \BibitemOpen
  \bibfield  {author} {\bibinfo {author} {\bibfnamefont {Albert-László}\
  \bibnamefont {Barabási}}, \bibinfo {author} {\bibfnamefont {Réka}\
  \bibnamefont {Albert}}, \ and\ \bibinfo {author} {\bibfnamefont {Hawoong}\
  \bibnamefont {Jeong}},\ }\bibfield  {title} {\enquote {\bibinfo {title}
  {Scale-free characteristics of random networks: the topology of the
  world-wide web},}\ }\href@noop {} {\bibfield  {journal} {\bibinfo  {journal}
  {Physica A: Statistical Mechanics and its Applications}\ }\textbf {\bibinfo
  {volume} {281}},\ \bibinfo {pages} {69--77} (\bibinfo {year}
  {2000})}\BibitemShut {NoStop}%
\bibitem [{\citenamefont {Bose}(1963)}]{SRG}%
  \BibitemOpen
  \bibfield  {author} {\bibinfo {author} {\bibfnamefont {R.~C.}\ \bibnamefont
  {Bose}},\ }\bibfield  {title} {\enquote {\bibinfo {title} {Strongly regular
  graphs, partial geometries and partially balanced designs.}}\ }\href@noop {}
  {\bibfield  {journal} {\bibinfo  {journal} {Pacific J. Math.}\ }\textbf
  {\bibinfo {volume} {13}},\ \bibinfo {pages} {389--419} (\bibinfo {year}
  {1963})}\BibitemShut {NoStop}%
\end{thebibliography}

%

\end{document}